\documentclass[a4paper,11pt]{article}
\usepackage{epsfig}
\usepackage[utf8]{inputenc}
\usepackage{cite}
\usepackage{hyperref}
\usepackage{epsfig}
\usepackage{amsmath}
\usepackage{amsfonts}
\usepackage{graphicx}
\usepackage{cite}
\usepackage{color}
\usepackage[dvipsnames]{xcolor}
\usepackage{multirow}
\usepackage{amssymb}
\usepackage{subcaption,graphicx}

\def\be{\begin{equation}}
\def\ee{\end{equation}}

\begin{document}
\title{{\bf  \LARGE Does the black hole shadow \\  probe the event horizon geometry?}}

 \author{
{\large Pedro V.P. Cunha}$^{1,2}$, \
{\large Carlos A. R. Herdeiro}$^{1}$, \
{\large Maria J. Rodriguez}$^{3,4}$,  \\
\\
$^{1}${\small Departamento de F\'\i sica da Universidade de Aveiro and } \\ {\small  Centre for Research and Development  in Mathematics and Applications (CIDMA),} \\ {\small    Campus de Santiago, 3810-183 Aveiro, Portugal}
 \\
 \\
$^{2}${\small Centro de Astrof\'isica e Gravitaç\~ao - CENTRA, Departamento de F\'isica,}\\ { \small Instituto Superior T\'ecnico - IST, Universidade de Lisboa - UL,}\\ {\small Av. Rovisco Pais 1, 1049-001, Lisboa, Portugal}
 \\
 \\
$^{3}${\small Max Planck for Gravitational Physics - Albert Einstein Institute,}\\ { \small Am M\"uhlenberg 1, Potsdam 14476, Germany} 
\\
\\ 
  $^{4}${\small  Department of Physics, Utah State University,}\\ {\small 4415 Old Main Hill, UT 84322, USA}\\
}


\maketitle

\begin{abstract}
There is an exciting prospect of obtaining the shadow of astrophysical black holes (BHs) in the near future with the Event Horizon Telescope. As a matter of principle, this justifies asking how much one can learn about the \textit{BH horizon itself} from such a measurement. Since the shadow is determined by a set of special photon orbits, rather than horizon properties, it is possible that different horizon geometries yield similar shadows. One may then ask how sensitive is the shadow to details of the horizon geometry? As a case study, we consider the double Schwarzschild BH and analyse the impact on the lensing and shadows of the conical singularity that holds the two BHs in equilibrium -- herein taken to be a strut along the symmetry axis in between the two BHs. Whereas the conical singularity induces a discontinuity of the scattering angle of photons, clearly visible in the lensing patterns along the direction of the strut's location, it produces no observable effect on the shadows, whose edges remain everywhere smooth. The latter feature is illustrated by examples including both equal and unequal mass BHs. This smoothness contrasts with the intrinsic geometry of the (spatial sections of the) horizon of these BHs, which is not smooth, and provides a sharp example on how BH shadows are insensitive to some  horizon geometry details. This observation, moreover, suggests that for the study of their shadows, this static double BH system may be an informative proxy for a dynamical binary. 
\end{abstract}

\bigskip

\tableofcontents

\section{Introduction}
In relativistic gravity, the propagation of light on curved spacetimes provides a basic probe of the background's properties. It reveals, of course, the causal structure of the spacetime; but it also unveils other relevant physical and phenomenological features. Indeed, the \textit{weak} lensing of light by the gravitational field of the Sun was the first successfully tested prediction of Einstein's general relativity~\cite{Dyson:1920cwa}.

\textit{Strong} lensing effects, on the other hand, can occur around very compact objects. In particular ultra-compact objects are, by definition, described by spacetimes that have bound photon orbits, dubbed \textit{fundamental photon orbits} (FPOs) in~\cite{Shipley:2016omi,Cunha:2017eoe,Cunha:2018acu}. This class of spacetimes includes black holes (BHs) but also horizonless compact objects - see $e.g.$~\cite{Cardoso:2014sna,Cardoso:2016rao,Cardoso:2016oxy,Cardoso:2017cqb}. For the Schwarzschild BH these orbits are all planar and circular; such special FPOs are known as \textit{light rings} (LRs).  For the Kerr BH, on the other hand, non-planar bound orbits arise, know as \textit{spherical photon orbits}~\cite{Teo2003}, in addition to LRs. For both Schwarzschild and Kerr all FPOs are unstable. But for a generic ultra-compact object, the set of FPOs can also include stable photon orbits -- see $e.g.$~\cite{Cardoso:2014sna,Cunha:2015yba,Shipley:2016omi,Dolan:2016bxj,Cunha:2016bjh,Cunha:2017wao,Cunha:2017qtt}. 

LRs (and other FPOs) generically  impact on the lensing properties of the spacetime. Unstable LRs yield divergences in the scattering of angle of photons -- see $e.g$~\cite{Cunha:2017wao}, whereas stable LRs can lead to chaotic scattering (and lensing)~\cite{Cunha:2015yba,Shipley:2016omi,Dolan:2016bxj,Cunha:2016bjh}. In the particular case of BHs, a set of unstable LRs (and other FPOs) determine the edge of the \textit{BH shadow}~\cite{Bardeen1973,Falcke:1999pj,ZAKHAROV2005479}, the absorption cross section at high frequencies under given observation conditions.  This shadow is a fingerprint of the BH spacetime, and an accurate measurement thereof could, in principle, pinpoint the precise type of BH that is being observed~\cite{Psaltis:2014mca,Bambi:2015rda,Johannsen:2015hib}. In practice, however, the light emitting astrophysical environment around the BH may cause degeneracies and make very different spacetimes potentially yield similar shadows - see $e.g.$~\cite{Vincent:2015xta,Mars:2017jkk} for examples.

It is therefore relevant to inquire how much BH shadows are a sensitive probe of the horizon geometry, even within ideal observation conditions. As a case study, we consider here the example of two interacting Schwarzschild BHs. Rather than a dynamical binary, whose spacetime geometry is time dependent and known only numerically, we study a toy model known as the double Schwarzschild (or Bach-Weyl~\cite{Bach:1922}) solution, a particular example of a Weyl solution~\cite{Weyl:1917gp}. This is an exact analytical solution of Einstein's equations describing a static, axially symmetric spacetime containing two Schwarzschild BHs at some distance. The BHs are kept apart by a conical singularity~\cite{Einstein:1936fp}, that plays the role of a strut (in our choice) preventing the two BHs from falling into each other. With this choice, the spacetime is asymptotically flat. This solution can be generalised to $N$ collinear Schwarzschild BHs, in what is known as the Israel-Khan solution~\cite{Israel:1964}.

The individual BHs in the double Schwarzschild solution have a deformed horizon geometry by virtue of the pressure exerted by the strut. This is easily visualised considering the isometric embedding of these horizons in Euclidean 3-space as discussed in~\cite{Costa:2000kf} and below. In particular the horizon is not everywhere smooth, possessing a singular point. As we shall show, however, the BH shadows are blind to this deformation, being smooth. In fact the (main) shadow presents similar features to that obtained in a dynamical binary~\cite{Bohn:2014xxa} wherein the individual BHs will certainly not present similar deformations of their horizon geometry. This example, albeit academic, shows clearly that BH shadows are not a faithful probe of the horizon geometry.\footnote{An earlier study of shadows in spacetimes with conical singularities can be found in~\cite{Grenzebach:2015oea}.} 

The insensitivity of the shadow to the conical singularity, does not mean the latter is irrelevant for the lensing. Rather, a lensing signature of the conical singularity appears as a discontinuity in the scattering angle for neighbouring null geodesics that circumvent the conical singularity from either side, giving rise to a clearly detectable pattern in the lensing; but not in the shadows. We will argue that due to the cylindrical symmetry in the problem, of both the metric and the spatial part of the FPOs, the conical singularity will produce no net effect in the azymuthal $\varphi$-direction.

This paper is organised as follows. In Section~\ref{section2} we describe the double Schwarzschild solution and present its shadows and lensing. In particular we compare the shadows obtained with those of a dynamical binary system and the horizon geometry of the BHs in the double Schwarzschild solution, presented in terms of embedding diagrams. In Section~\ref{section3}, the role of FPOs is discussed on an emitting star's outline in Schwarzschild and Kerr. Closing remarks are presented in Section~\ref{section4}.

\section{Shadows in the double Schwarzschild BH solution}
\label{section2}
In this section we review the double Schwarzschild BH solution and compute its shadow. In particular, the effect of the conical singularity  is shown to have no significant effect on the shadow. This observation is contrasted with the behaviour of the intrinsic horizon geometry, analysed through embedding diagrams, as discussed toward the end of this section.

\subsection{Double Schwarzschild solution review}

The double Schwarzschild BH is a static Weyl solution with axial-symmetry, featuring two non-rotating, neutral BHs supported in equilibrium by a conical singularity which can be chosen to take the form of either two strings or one strut (see $e.g.$~\cite{PhysRevD.80.104036}). The metric can be reduced to the form:
\[ds^2=-e^{2U}dt^2 + e^{-2U}\left(e^{2K}\left[d\rho^2 +dz^2\right] +\rho^2 d\varphi^2\right),\]
where $t,\varphi$ are connected respectively to staticity and axial-symmetry. Due to these symmetries, both $U(\rho,z)$ and $K(\rho,z)$ are only functions of $\rho,z$. With this ansatz, it is well known that the vacuum Einstein equations reduce to
\be
\Delta_{\mathbb{E}^3}U(\rho,z)= 0 \ , \label{harmonic} \ee
where the operator  $\Delta_{\mathbb{E}^3}$ represents the  Laplacian in an auxiliary Euclidean 3-space with line element
\[ ds^2_{\mathbb{E}^3}=d\rho^2+\rho^2d\varphi^2+dz^2 \ , \]
and
\be
\frac{\partial K}{\partial \rho}=\rho\left[\left(\frac{\partial U}{\partial \rho}\right)^2-\left(\frac{\partial U}{\partial z}\right)^2\right] \ , \ \ \ \ \frac{\partial K}{\partial z}=2\rho\frac{\partial U}{\partial \rho}\frac{\partial U}{\partial z}\ . \label{k}\ee
The problem of finding an exact solution of Einstein's equations then reduces to the \textit{linear} equation~\eqref{harmonic}, which has the interpretation of a Newtonian problem with some mass distribution along the $z$-axis in the auxiliary 3-space. Once such distribution is fixed, the potential $U$ is determined and the functions $K$ are obtained by solving the line integrals~\eqref{k}.
In this parametrisation, the Schwarzschild solution of mass $M$ corresponds to choosing the source of~\eqref{harmonic} to be a zero thickness mass rod along the $z$ axis, with $z$-coordinate length $2M$ and linear mass density $1/2$. Taking two such rods, on the other hand, one obtains the double Schwarzschild (a.k.a. Bach-Weyl or 2-centre Israel-Khan) solution. In this case, explicitly, the functions $U,K$ satisfy
\[e^{2U}=\frac{\gamma_1 \gamma_3}{\gamma_2 \gamma_4}, \qquad e^{2K}=\frac{Y_{43}Y_{21}Y_{41}Y_{32}}{4Y_{42}Y_{31}R_1R_2R_3R_4},\]
where
\[R_k\equiv \sqrt{\rho^2 +(z-a_k)^2},\qquad \gamma_k \equiv R_k+a_k-z,\qquad Y_{ij} \equiv R_iR_j + (z-a_i)(z-a_j)+ \rho^2.\]

\begin{figure}[h!]
\begin{center}
\includegraphics[width=0.22\textwidth]{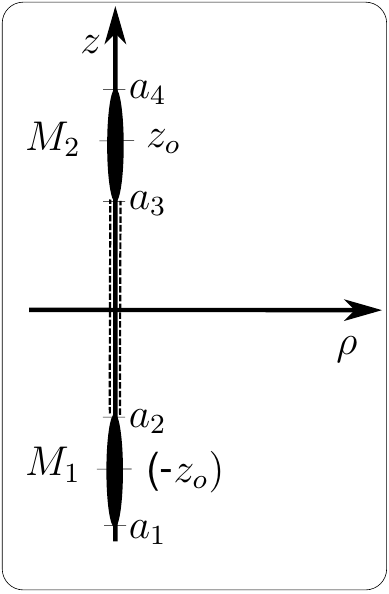}
\caption{\small Schematic representation of the double Schwarzschild BH system with the used parametrisation. The solid black rods along the $z$-direction represent each BH horizon while the dashed line in between these rods correspond to the conical singularity. The two BH masses (computed as Komar integrals on each horizon) are $M_1,M_2$. The  three independent parameters of the solution can be taken, for instance, as the mass difference $\epsilon\equiv M_2-M_1$, the total (ADM) mass $M=M_1+M_2$ and the distance parameter $z_o$.}
\label{setup}
\end{center}
\end{figure}

In this coordinate system, the two BH horizons are line segments on the $z$-axis. The parameters $a_k$ define the positions of the horizons and masses of the BHs, with $a_1<a_2\leqslant a_3<a_4$. In particular, one of the horizons exists along the interval $z\in[a_1,a_2]$ and the other one along $z\in[a_3,a_4]$, with a conical singularity strut in between (see Fig.~\ref{setup}). 
Up to a $z$ origin shift, the most general parametrisation is provided by: 
\[a_1=-\frac{1}{2}(M-\epsilon) -z_o,\qquad a_2=\frac{1}{2}(M-\epsilon) -z_o,\]
\[a_3=-\frac{1}{2}(M+\epsilon) +z_o,\qquad a_4=\frac{1}{2}(M+\epsilon) +z_o,\]
where $M$ denotes the ADM mass and $\epsilon=M_2-M_1$ the mass difference between the BHs. Both $M_2,M_1$ are determined via Komar integrals, with the former (latter) corresponding to the BH with larger (smaller) $z$. Additionally, the value of $\epsilon$ can be related to the mass ratio $\mu=M_2/M_1$ between the BHs via $\epsilon=M(\mu-1)/(1+\mu)$.
The position of both BHs on the $z$-axis is also set by the parameter $z_o$, with the latter being related to the BH coordinate distance $L=(a_3-a_2)/2$ via $L=z_o-M/2$. The lower limit of $z_o$ is bounded by the condition $L=0$, yielding the allowed range $z_o\in[M/2,+\infty[$. Notice that for $\epsilon=0$, the solution has a $\mathbb{Z}_2$ reflection symmetry on the equatorial plane $(z=0)$.

\subsection{Shadows in double BH solution}

We wish to study the shadows of the double Schwarzschild solution reviewed in the previous section. Consider an observer with a local sky $O$, a manifold with $S^2$ topology. Each point in $O$ defines a direction of observation of an incoming null geodesic. The \textit{shadow} is the set of points in $O$ that leads to the infall of the associated null geodesics into the event horizon (which in our case has two disconnected components), when propagated backwards in time \cite{Cunha:2018acu}. Astrophysically, the shadow is then the region in the local sky that would receive light from the event horizon, but since the latter is not a source of radiation, at least classically, the shadow actually corresponds to a \textit{lack}\footnote{A sharp decrease in luminosity is still expected if light sources exist in between the observer and the event horizon.} of radiation, hence its name.

$O$ can be represented by a 2D \textit{observation image}, in which the pixel color encodes the endpoint of the associated geodesic (when propagated backwards in time). In particular, pixels that are part of the shadow are defined \textit{black}, whereas the remainder is given the endpoint color on a large far-away sphere $\mathcal{N}$, surrounding both the observer and the BHs (see~\cite{Cunha:2018acu}). More formally, the 2D image is a map between points in $O$ and the sphere $\mathcal{N}$, with the exception of the shadow, which is a map from $O$ to one of the BHs. The vertical (horizontal) axis of the image represents the latitude (longitude) angle of the local sky $O$, with both axis intersecting at the image center (see top left image of Fig.~\ref{th_change}), which is always pointing to the origin of the coordinated system ($\rho=z=0$).

\begin{figure}[h!]
\begin{center}
\includegraphics[width=0.25\textwidth]{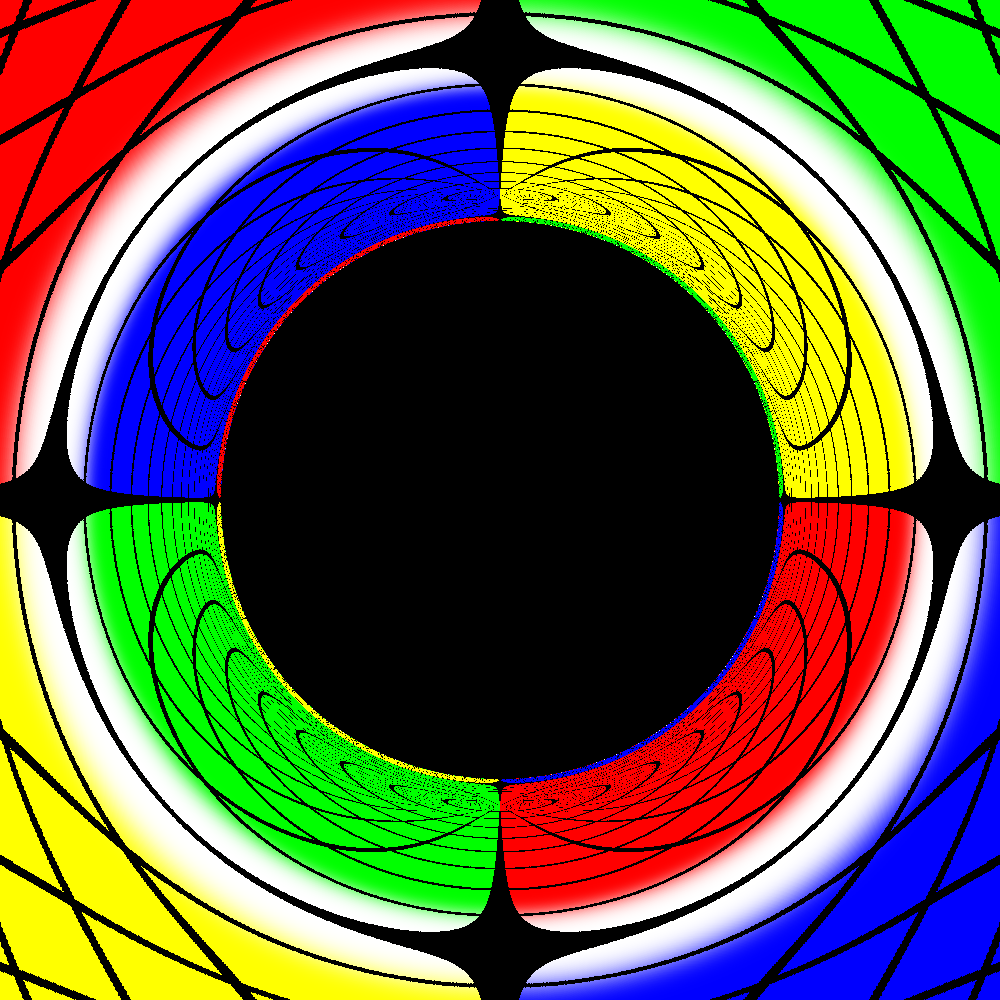}
\includegraphics[width=0.25\textwidth]{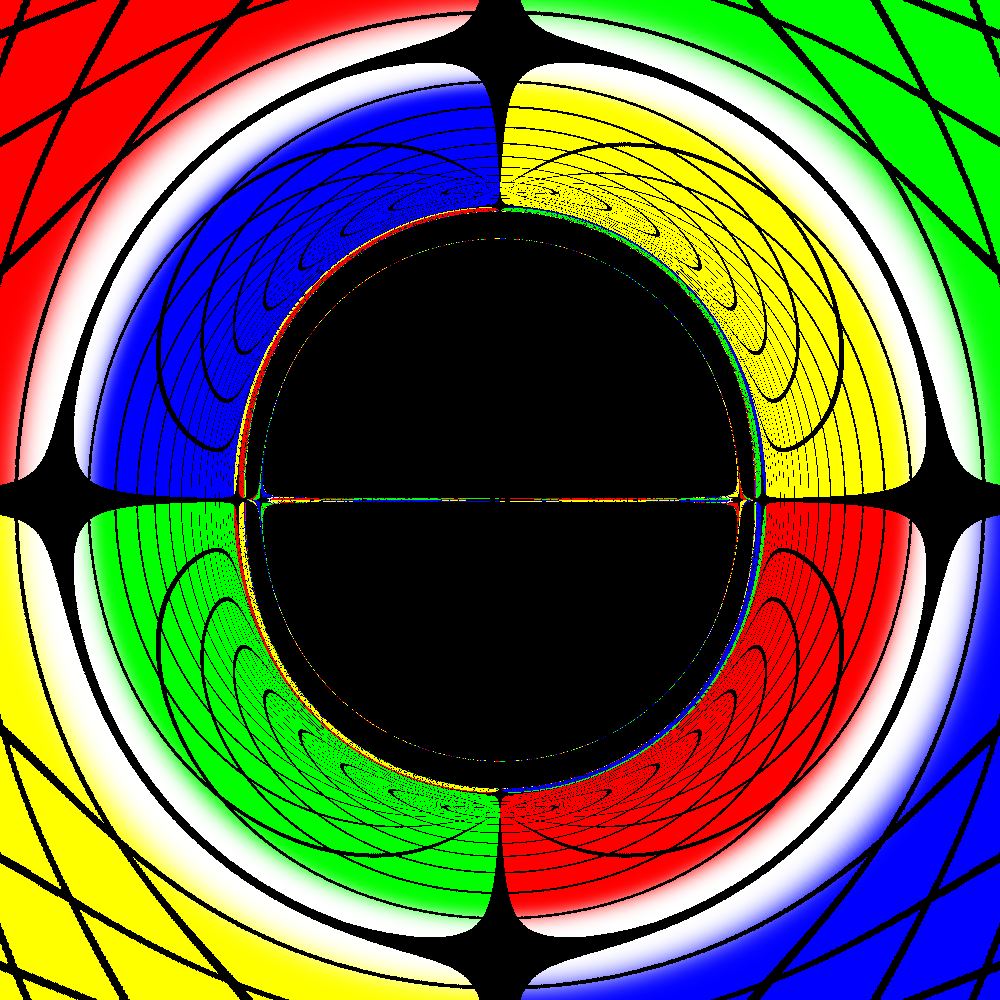}
\includegraphics[width=0.25\textwidth]{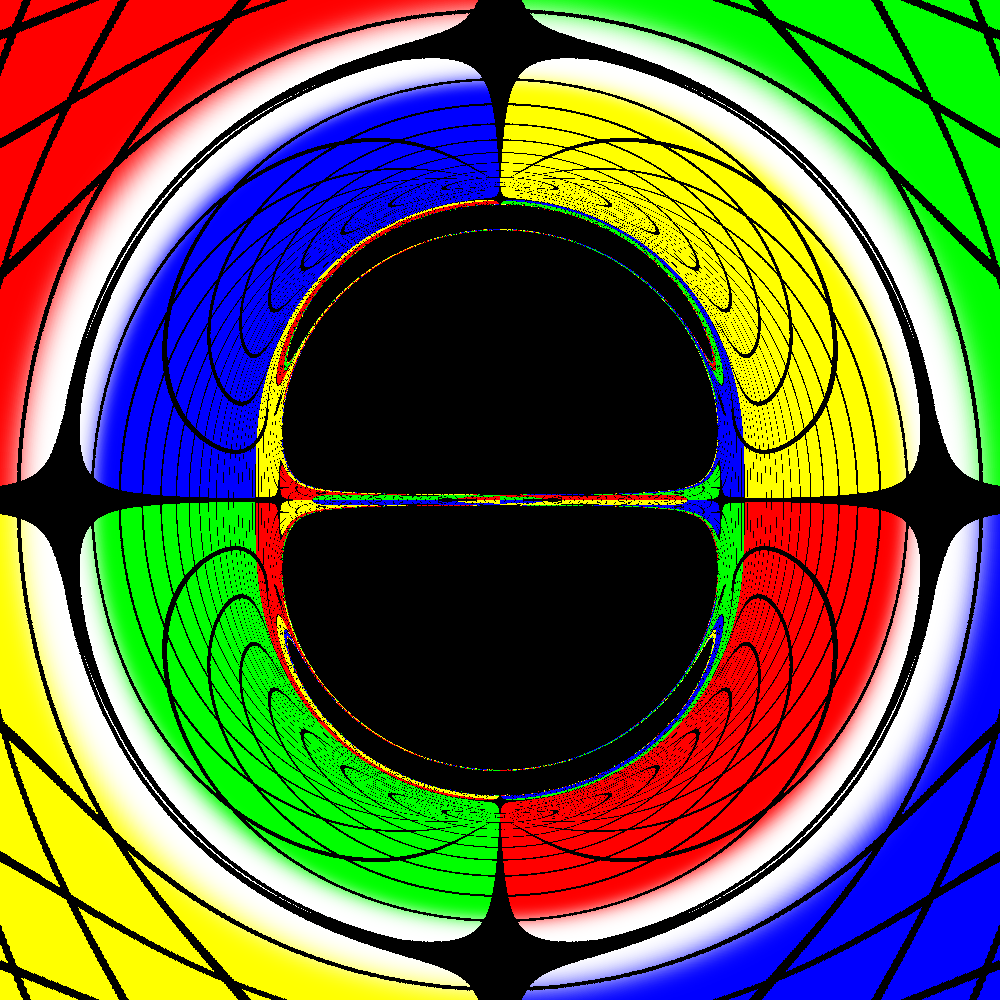}\\
\includegraphics[width=0.25\textwidth]{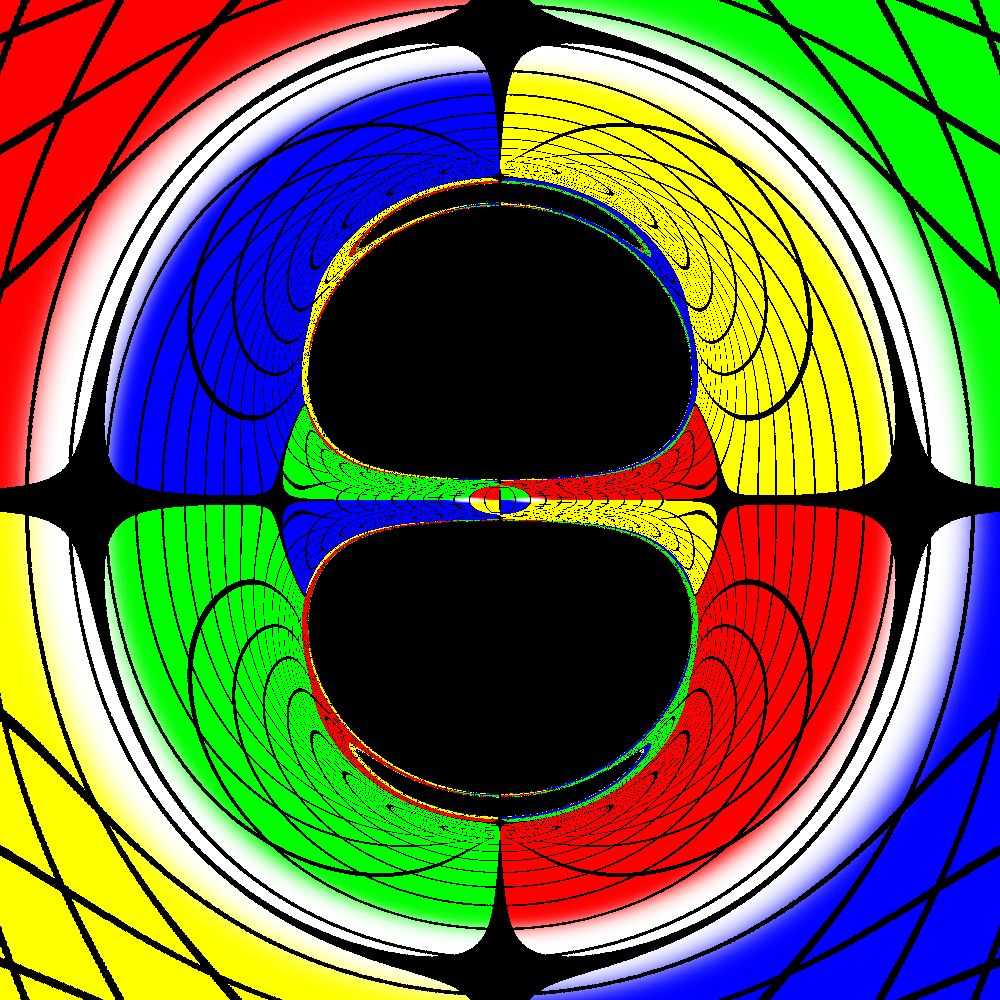}
\includegraphics[width=0.25\textwidth]{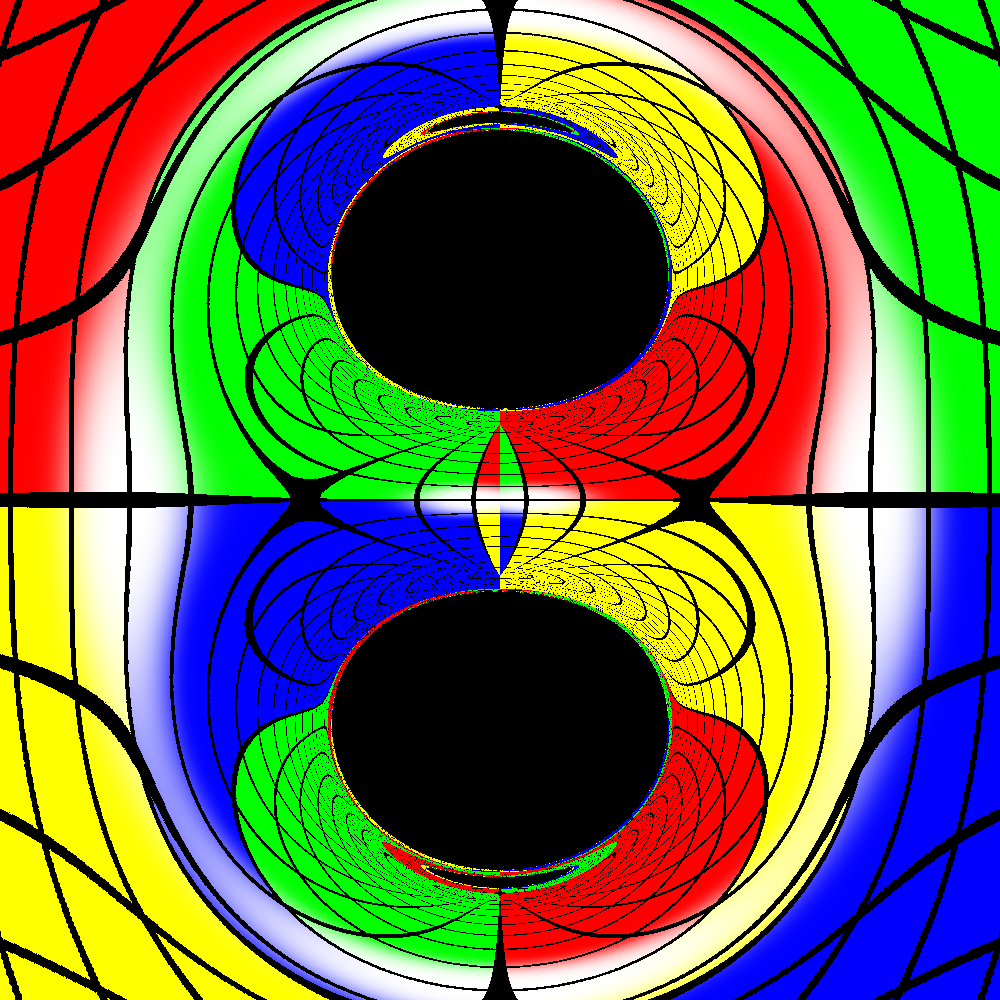}
\includegraphics[width=0.25\textwidth]{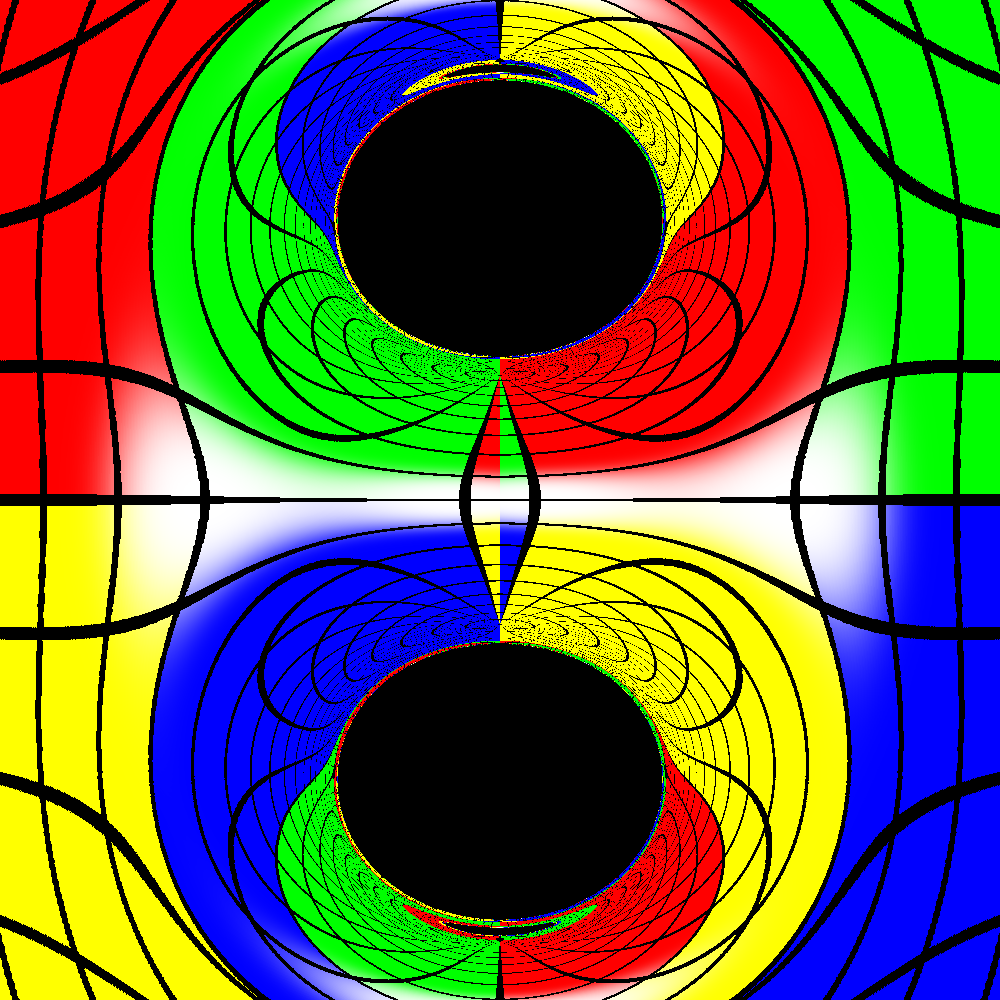}
\caption{\small Shadows of the double Schwarzschild BH solution with equal masses ($\epsilon=0$) and different BH distances, corresponding to a value $z_o/M$ of: (from left to right) (top) 0.5, 0.81, 1; (bottom): 1.5, 3 , 4.}
\label{zo_change}
\end{center}
\end{figure}

Following previous work~\cite{Bohn:2014xxa,Cunha:2015yba,Cunha:2016bjh,Cunha:2017eoe}, the sphere $\mathcal{N}$ is given four color quadrants, all imprinted with a regular grid. The observer is placed at a constant coordinate $r_o\equiv \sqrt{\rho^2+z^2}$, fixed by the perimetral radius $\mathcal{R}\equiv\sqrt{g_{\varphi\varphi}}=15M$ for $z=0$. The observation angle $\theta_o=\arccos(z/r_o)$ is $\pi/2$, unless otherwise specified. A white dot is also added to the sphere $\mathcal{N}$ in order to appear in the image center when $\theta_o=\pi/2$ in flat spacetime. 

Consider first the observation images in Fig.~\ref{zo_change} for the case of same mass BHs with $\epsilon=0$. Starting with $z_o=M/2$, the (standard) Schwarzschild shadow can be seen in the leftmost image of the top row, which is completely circular due to spherical symmetry. The white dot is stretched into a (white) ring, disclosing the location of an \textit{Einstein ring}. Inside this ring the entire sphere $\mathcal{N}$ is mapped an infinite number of times, as we approach the edge of the shadow. As we increase the distance between the two BHs, increasing the value of $z_o/M$, the shadow is broken into two large disconnected parts, each associated to a different BH. However, smaller shadows also exist, \textit{eyebrows}~\cite{Yumoto:2012kz}, which (heuristically) correspond to the lensing of a given BH's shadow by the other BH~\cite{Yumoto:2012kz,Bohn:2014xxa,Abdolrahimi:2015rua,Cunha:2016bjh,Shipley:2016omi}. It is worth mentioning, that the shadows presented in Fig.~\ref{zo_change} have two reflection symmetries; one is along the vertical axis and it is associated to the spacetime invariance $\varphi\to-\varphi$, whereas the other one is along the horizontal axis and it is inherited from observations at the $\mathbb{Z}_2$ symmetry plane $z=0$.

In order to assess the influence of a different angle of observation $\theta_o$, we also generated the corresponding shadows in Fig.~\ref{th_change}, wherein the colored sphere $\mathcal{N}$ was painted white for clarity. To further illustrate the image axis, these are displayed as dotted lines in the leftmost top image of Fig.~\ref{th_change}, with both axis intersecting each other in the image center. As one moves away from the $\mathbb{Z}_2$ plane $z=0$, the shadow is no longer symmetric along the the horizontal axis (the vertical reflection symmetry still holds however). Nevertheless, as the observer approaches the $z$-axis, the shadow becomes increasingly circular as a result of the spacetime axial symmetry, but it never becomes simply connected: it is rather deformed into a \textit{Saturn-like} shape (see also \cite{Cunha:2016bjh} for a similar effect).

\begin{figure}[h!]
\begin{center}
\includegraphics[width=0.4\textwidth]{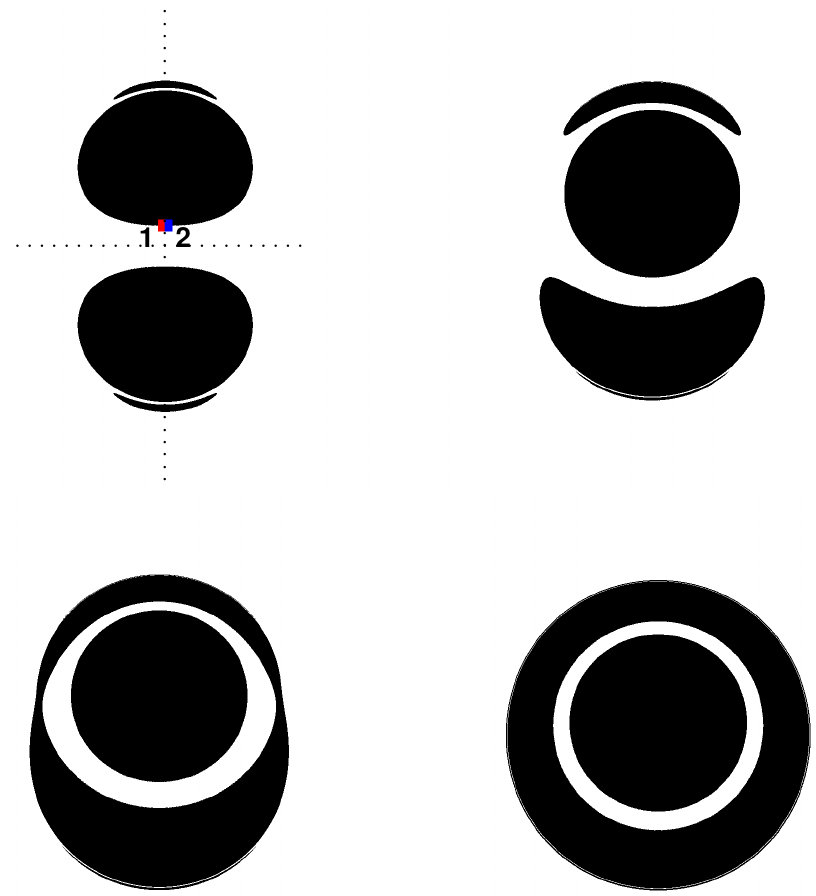}
\caption{\small Shadows of the double equal mass ($\epsilon=0$) Schwarzschild BH system, separation $z_o=2M$ and an observation angle $\theta_o$ of: (from left to right) (top) $90^\circ$, $40^\circ$; (bottom): 30$^\circ$, 10$^\circ$. For clarity, here the colored sphere $\mathcal{N}$ was painted white.}
\label{th_change}
\end{center}
\end{figure}

\begin{figure}[h!]
\begin{center}
\includegraphics[width=0.21\textwidth]{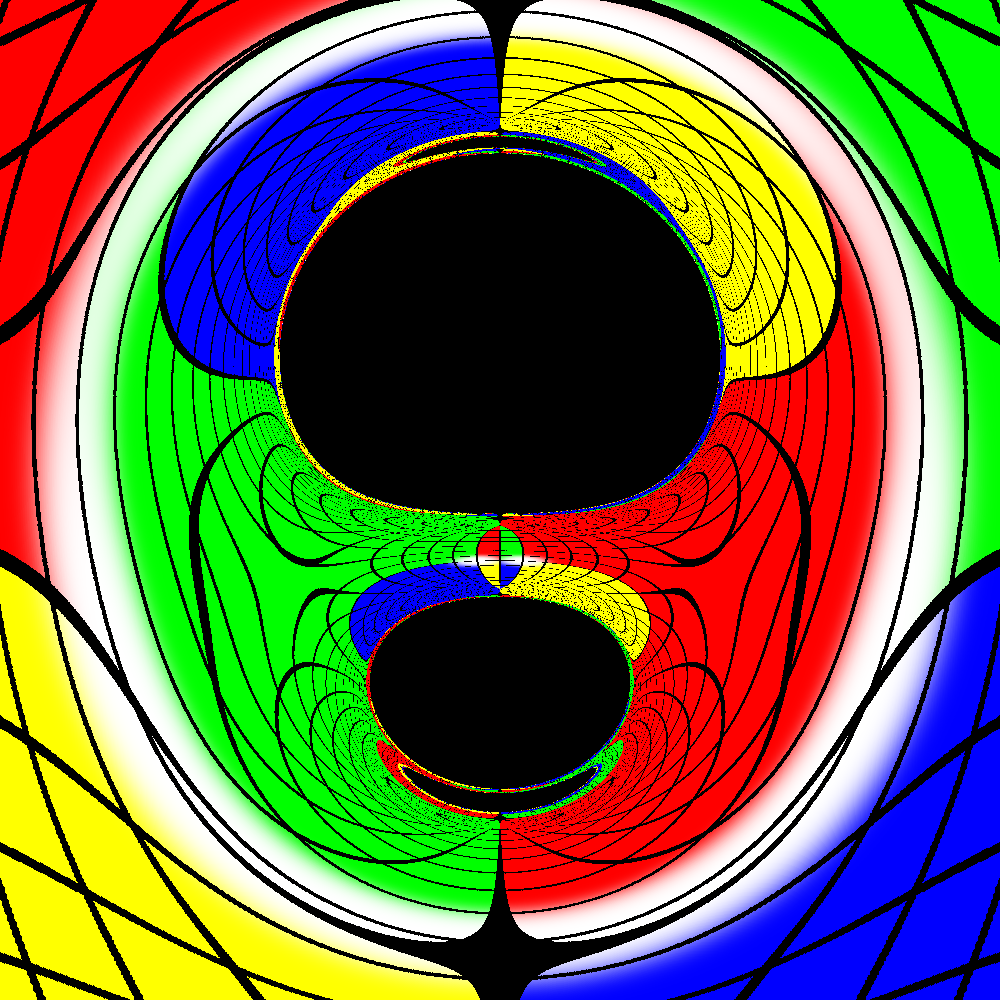}
\includegraphics[width=0.21\textwidth]{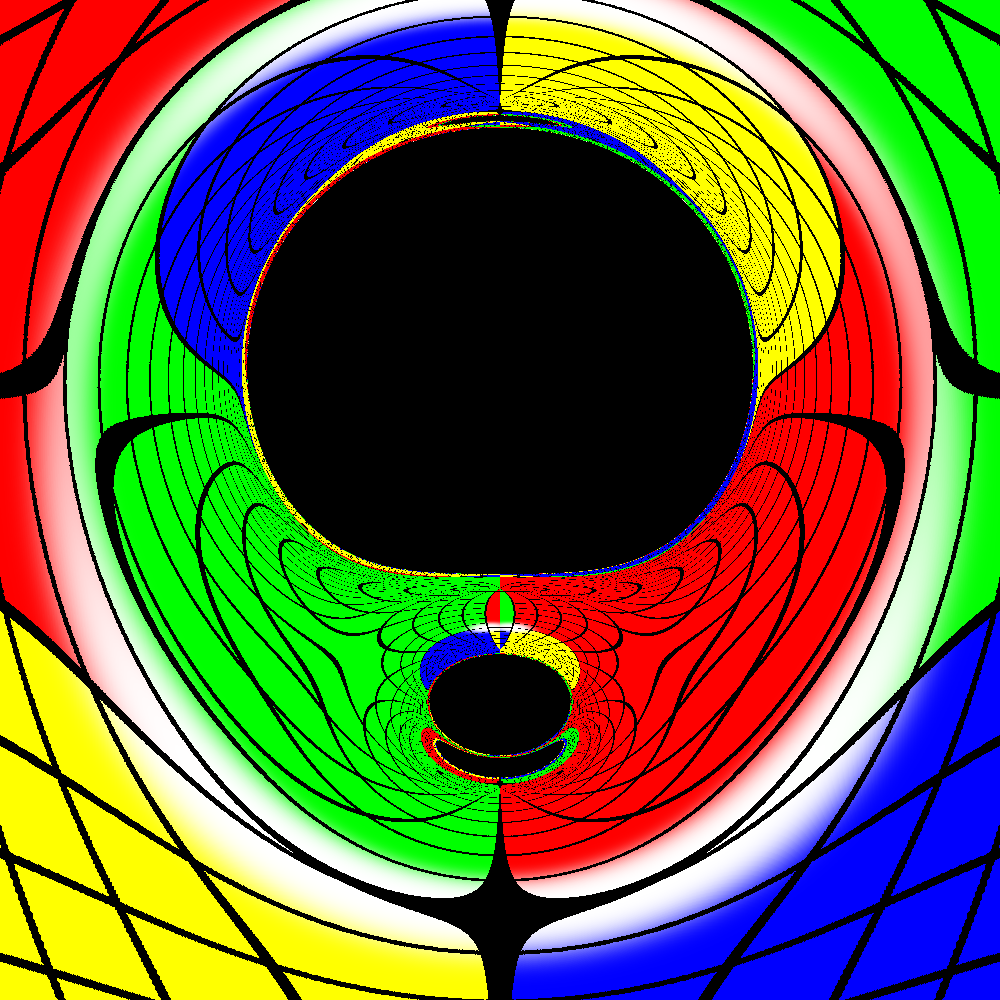}\\
\includegraphics[width=0.21\textwidth]{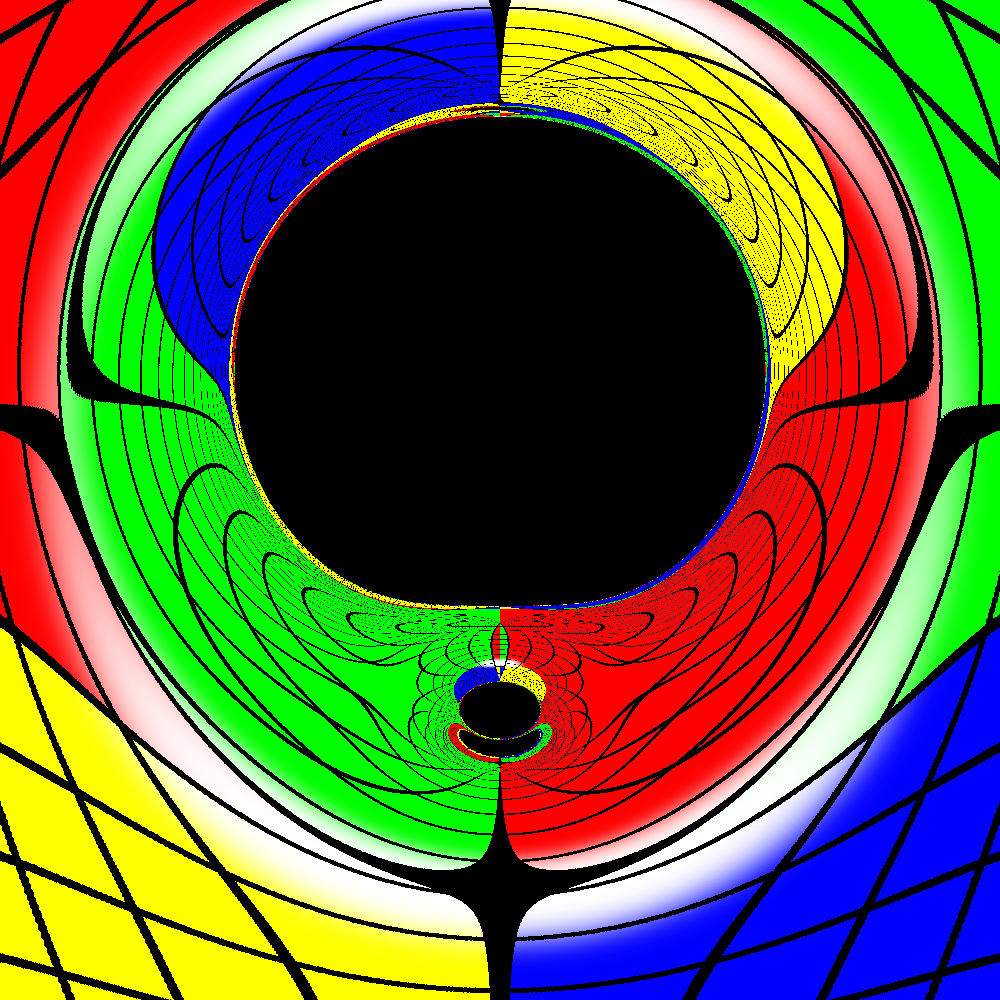}
\includegraphics[width=0.21\textwidth]{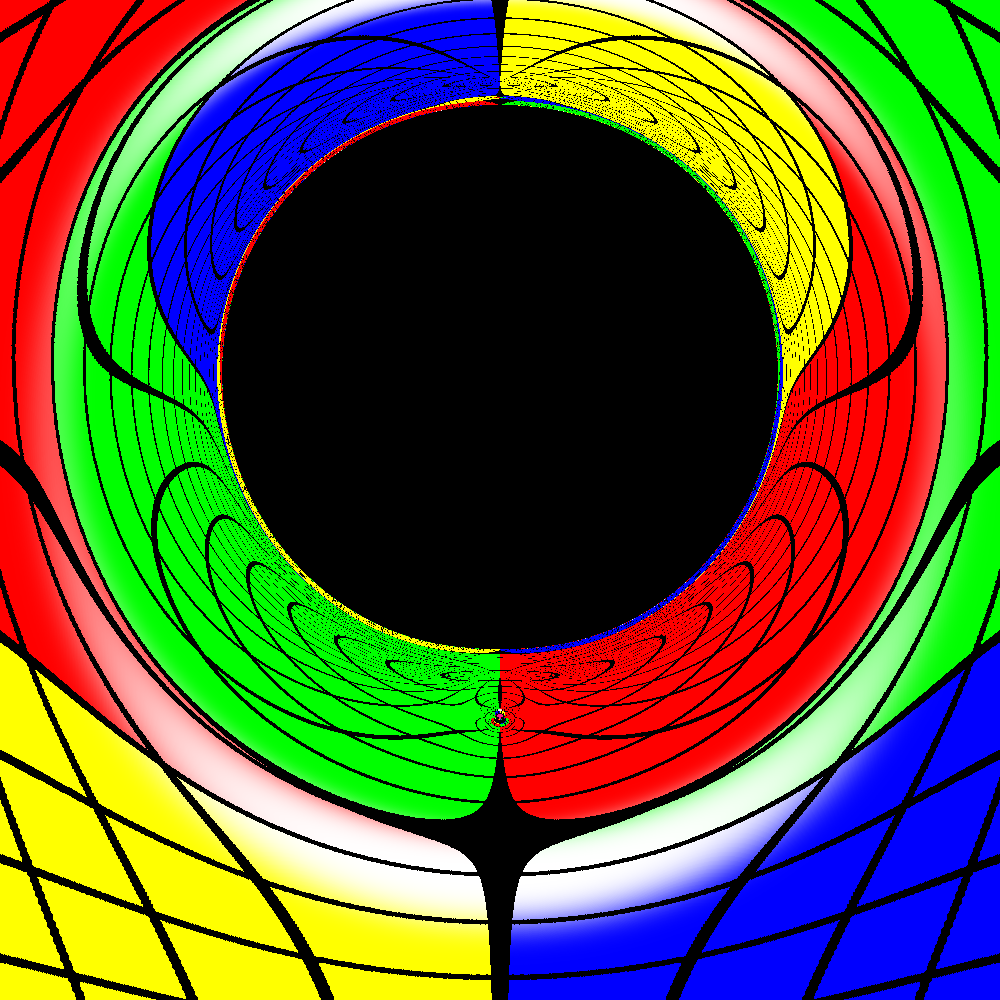}
\caption{\small Shadows of the double Schwarzschild BH system with separation $z_o=2M$ and a mass ratio $\mu$ of: (from left to right) (top) 2, 5; (bottom): 10, 100.}
\label{Mratio_change}
\end{center}
\end{figure}

For BHs with different masses, the $\mathbb{Z}_2$ reflection symmetry at $z=0$ is broken. In the discussion section we will comment on a potential implication of this observation. The shadows can be found in Fig.~\ref{Mratio_change}, wherein different mass ratios $\mu$ are analysed. In the limit of large $\mu$ the lensing becomes that of a single Schwarzschild BH but with $\theta_o\neq \pi/2$.

\subsection{Insensitivity to the conical singularity}

\begin{figure}[h!]
\begin{center}
\includegraphics[width=0.45\textwidth]{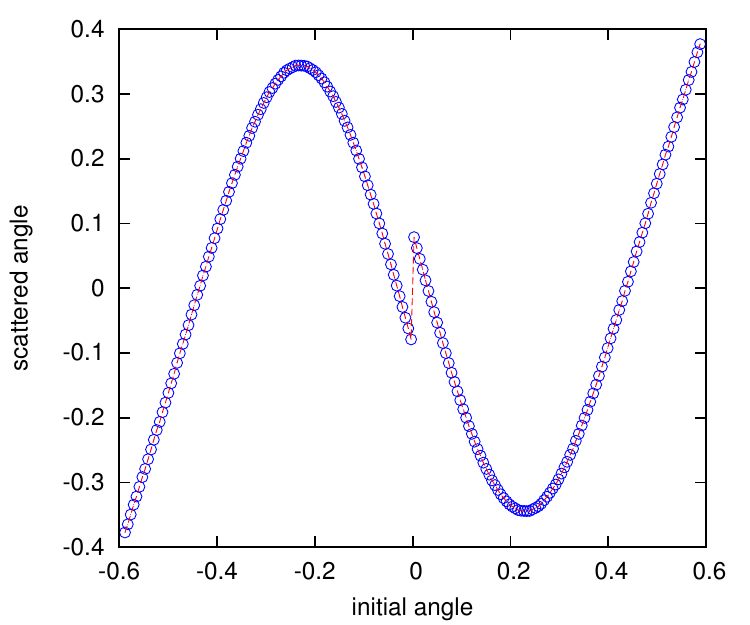}
\caption{\small Scattered angle, $i.e.$ coordinate $\varphi$ in $\mathcal{N}$, as a function of the observation angle along the horizontal image axis ($O$). The origin of the initial angle corresponds to the image center. The solution has $z_o=3M$, $\epsilon=0$. The jump $\delta\varphi$ in the middle corresponds to the effect of the conical singularity.}
\label{conical}
\end{center}
\end{figure}

We now turn to the analysis of the effect on the shadows in the double Schwarzschild solution and, show that due to the cylindrical symmetries in these solutions there is no effect of the conical singularity on their shadows. The conical singularity produces a subtle discontinuity in the geodesic scattering, perceptible by a sharp color transition in the vertical axis in between the two shadows (see Fig.~\ref{zo_change}). However, this effect becomes clearer if the scattered angle in $\mathcal{N}$ is plotted against the initial angle in $O$ (see Fig.~\ref{conical}).
\begin{figure}[h!]
\begin{center}
\includegraphics[width=0.45\textwidth]{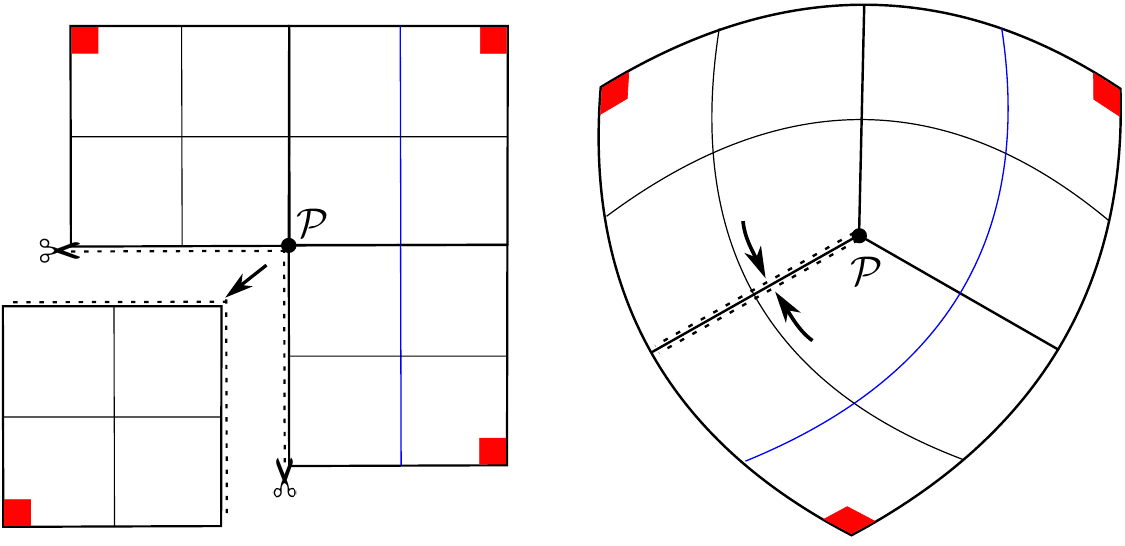}
\caption{\small By removing a section of a flat sheet of paper and gluing together the cutting edges, one creates an angular deficit $\alpha$ at point $\mathcal{P}$ (in the illustration $\alpha=-\pi/2$). This point is a conical singularity, leading to an angular deviation $\alpha=\delta\varphi$ between geodesics that circumvent $\mathcal{P}$ from either side.}
\label{conical-drawing}
\end{center}
\end{figure}
The conical singularity produces an angular difference $\delta\varphi$ between geodesics that circumvent the conical axis from either side, which can be computed analytically
\begin{eqnarray}
\delta\varphi &\equiv& 2\pi \,\lim_{\rho\to0} \left(e^{-K}-1\right), \quad a_2<z<a_3,\nonumber \\
&=& 2\pi\left(\frac{M^2-\epsilon^2}{4z_o^2-M^2}\right) \,
\end{eqnarray}
Notice that $\delta\varphi$ vanishes as $z_o\to\infty$, $i.e.$ as the BHs become infinitely far away from one another, or when $\epsilon=M$, which corresponds to the single BH limit. We find that the value of $\delta\varphi$ computed analytically is consistent with the numerical displacement of the scattering angle in Fig.~\ref{conical}.

In order to further illustrate this effect, while providing some additional physical intuition, consider the simple procedure depicted in Fig.~\ref{conical-drawing}. Starting from a flat sheet of paper, one can make two straight cuts that intersect with an angle $\alpha$ at some point $\mathcal{P}$, discarding the piece that detaches afterwards. By gluing the cutting edges together, one creates a conical surface with a cusp in $\mathcal{P}$, hence forming a \textit{conical} singularity with respect to the surface. By construction, any simply connected region not including $\mathcal{P}$ is flat, despite the global curvature introduced by the conical singularity. This is well illustrated on the right image of Fig.~\ref{conical-drawing}, wherein the outermost triangle (of geodesics) has three red angles that sum $270^\circ>180^\circ$. More generically, a triangle of geodesics encircling $\mathcal{P}$ has internal angles summing $180^\circ - \alpha$, whereas triangles not encircling $\mathcal{P}$ still sum $180^\circ$.

In addition, the conical singularity leads to an angular deflection of nearby geodesics as illustrated by the blue line in Fig.~\ref{conical-drawing}. In the illustrated case, $\alpha<0$ and the conical singularity is \textit{attractive}. In contrast, $\alpha>0$ would lead to a \textit{repulsive} $\mathcal{P}$, corresponding to an angle \textit{excess} rather than a deficit; this is actually the case of the conical singularity in the double Schwarzschild solution that we are analysing.
From Fig.~\ref{conical-drawing} it is also clear that a geodesic that barely skims $\mathcal{P}$ is deflected by $\alpha/2$, leading to an angular deviation $\delta\varphi=\alpha$ between geodesics that circumvent $\mathcal{P}$ from either side.

\begin{figure}[h!]
\begin{center}
\includegraphics[width=0.3\textwidth]{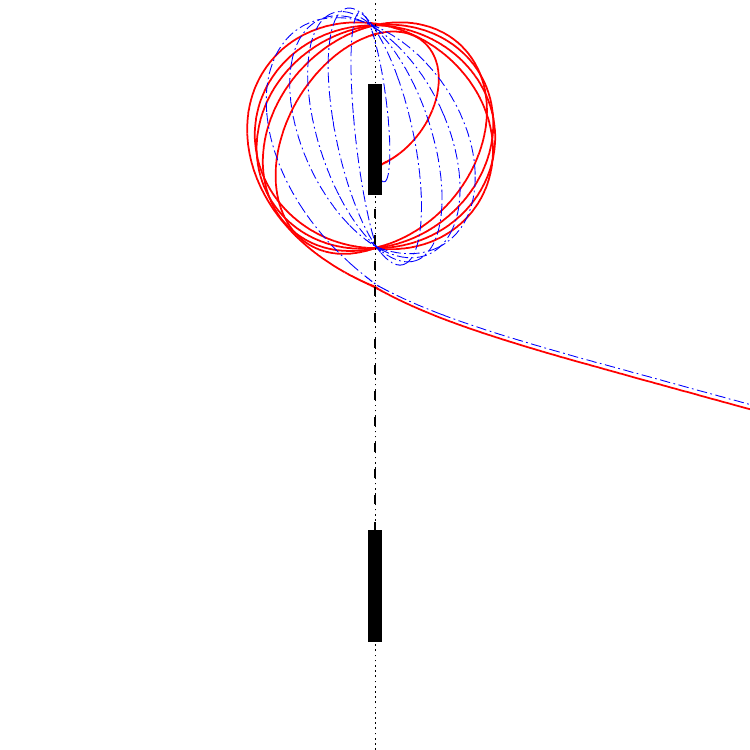}\includegraphics[width=0.24\textwidth]{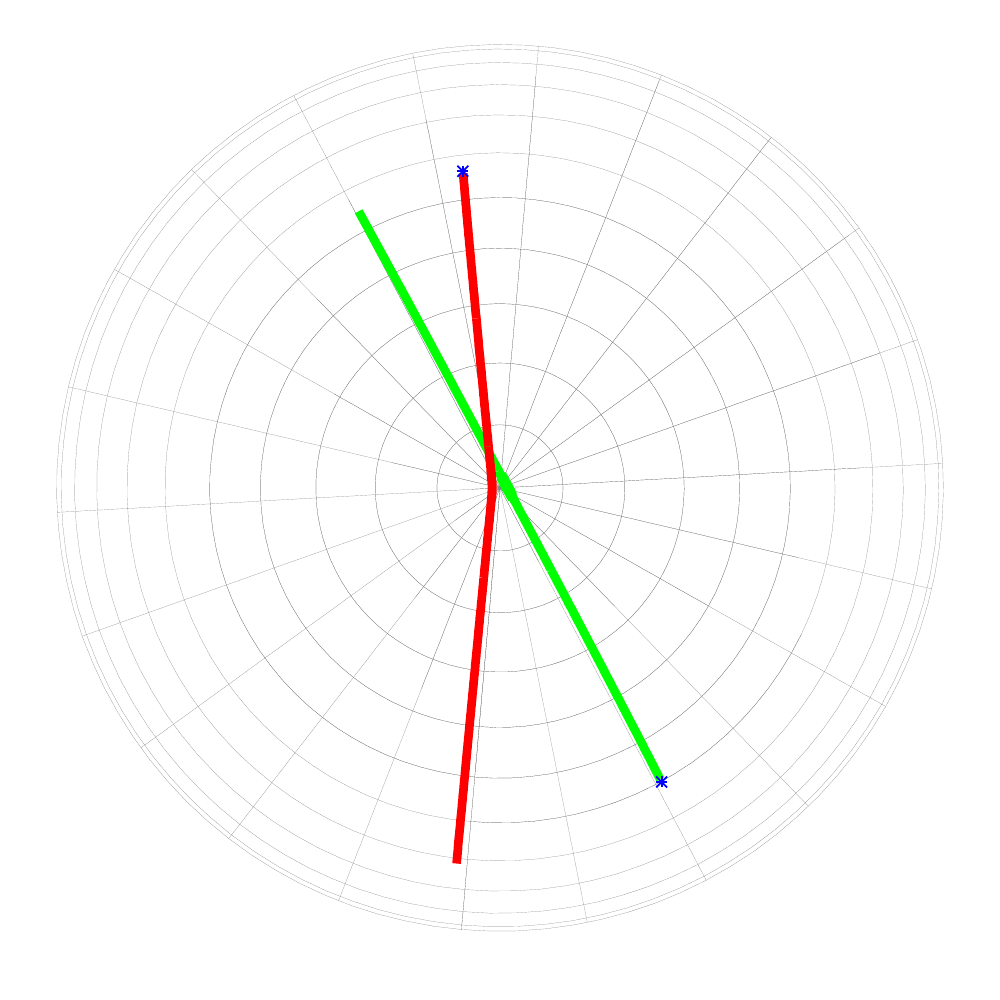}
\caption{\small {\it Left:} representation of the geodesics associated to two points in the shadow edge ({\bf 1},{\,\bf 2} in the top left of Fig.~\ref{th_change}), with both geodesics approaching an FPO surface, which has a spherical-like profile; the coordinates ($\rho,z,\varphi$) were represented as if they were cylindrical, with each BH being represented by a black line segment along the dashed $z$-axis. {\it Right:} Representation as seen from the $z$-axis of two geodesics (red and green) on the FPO surface that just skim the axis, with the polar mesh representing $(\rho,\varphi)$. The red (green) geodesic approaches the $z$-axis from below (above) the uppermost BH, with the geodesic suffering (not suffering) an angular deflection.}
\label{FPO}
\end{center}
\end{figure}

Surprisingly, as can be easily observed from the previous images, the conical singularity has no clear effect on the shadow edge, in sharp contrast with the jump of Fig.~\ref{conical}. Within the numerical accuracy, the shadows always appear to be smooth and without cusps. However, this can be expected, since the edge of the shadow corresponds to geodesics asymptotically approaching a special class of orbits: FPOs~\cite{Cunha:2017eoe}. The spatial part of FPOs typically exists on a 2-surface with cylindrical topology, invariant under the action of the Killing vector $\partial_\varphi$. Consequently, a deflection $\delta\varphi/2$ will produce no net effect for geodesics approaching an FPO.

To have a clearer depiction of these findings we represent in Fig.~\ref{FPO} ({\it left}) two geodesics, colored in red and blue. These geodesics correspond to two points ({\bf 1} and {\bf 2}) very close to the shadow edge, labeled by the respective color in the top left image of Fig.~\ref{th_change}. These shadow points exist very close to - and on both sides - of the vertical image axis, leading to geodesics that approach a spherical-like surface ({\it i.e.}\,an FPO), while barely skimming the conical singularity. This spherical-like FPO surface has two (very small) openings close to the $z$-axis, respectively above and below the uppermost BH. Since the conical singularity only exists in between the BHs, one can expect an angular deflection $\delta\varphi/2$ for geodesics that comes very close to the lower FPO axis opening (but not the upper one). Indeed, this is the case, as illustrated in the right image of Fig.~\ref{FPO}. Nevertheless, despite this deflection, the conical singularity appears to have no significant effect at the level of the FPO structure, which is what is critical for the shadow edge (and its smoothness).

\subsection{Horizon geometry embedding, shadows and dynamical binary BH}

\begin{figure}[h!]
\begin{center}
\includegraphics[width=0.5\textwidth]{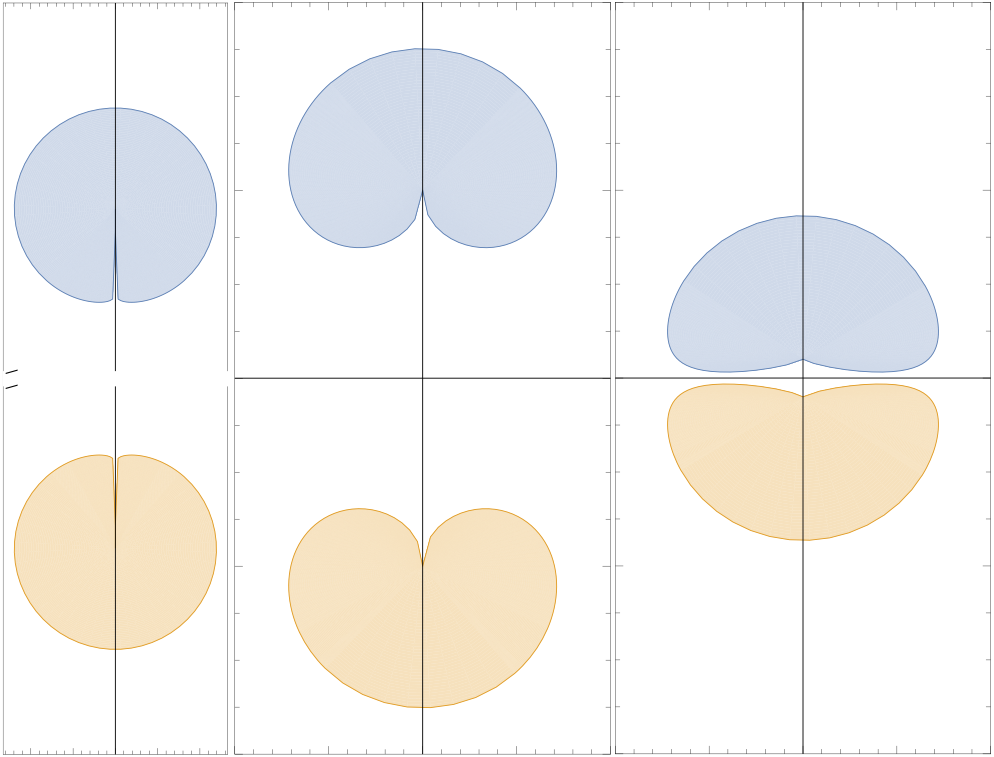}
\caption{\small Global embedding in Euclidean 3-space - in cartesian $(x,y)$-coordinates -  of the individual horizons of the double Schwarzschild solution $(\epsilon=0)$, with parameters ({\it from left to right}) $z_o=\{3,\,0.75,\,0.525\}M$. These images are in contrast with the shadows in that the latter does not present any cusps due to the conical singularity (see Fig. \ref{zo_change})}. 
\label{Horizons}
\end{center}
\end{figure}

In sharp contrast to the shadow, the conical singularity has an important effect on the intrinsic geometry of the individual horizons in the double Schwarzschild solution. This effect can be visualised by performing a global embedding of the individual (spatial sections of the) horizon in Euclidean 3-space. For the case of the double Schwarzschild solution (in contrast to the double Kerr solution~\cite{Costa:2009wj,Gibbons:2009qe}), such global embedding is always possible~\cite{Costa:2000kf} and the result is provided in Fig.~\ref{Horizons} for the equal mass double Schwarzschild solution for three different separations. Comparing Fig.~\ref{Horizons} (horizon geometry) with Fig. \ref{zo_change} (shadows) one can see that the shadow is blind to the (intuitive) sharp edge induced by the strut. Furthermore, the main shadows of the double Schwarzschild solution resemble the ones in a dynamical binary, which has no conical singularities. Indeed, comparing Fig.~\ref{zo_change} (say, the bottom left panel) with Fig. 5 in~\cite{Bohn:2014xxa} of a fully dynamical BH merger one observes a striking similarity in the main shadows, whereas the eyebrows in the latter are slightly displaced (likely) as a  consequence of the motion of the BHs in the dynamical binary. The overall lensing, on the other hand, presents some differences, most interestingly, the imprint of the conical singularity along the symmetry axis in the static two BH system, which is absent in the dynamical binary. Thus, we conclude that the shadows are totally blind to the conspicuous geometrical deformation of the horizon caused by the conical singularity.

\section{Emitting star surface in static and spinning BHs}
\label{section3}

In the previous section we computed the shadows of the double Schwarzschild BH solution and made a tentative general statement about the insensitivity of the shadow with respect to the detailed BH horizon geometry. In this section we provide support to this idea by considering the following academic exercise: an emitting (star) surface is placed in a Schwarzschild or Kerr spacetime at some radial function $R(\theta)$, in Boyer-Lindquist coordinates ($r,\theta$). As the star (mean) radius decreases and approaches the event horizon, how is its image changed?\\

\begin{figure}[htp] 
     \centering
    \begin{subfigure}{0.5\textwidth}
\centering
 \includegraphics[width=0.4\textwidth]{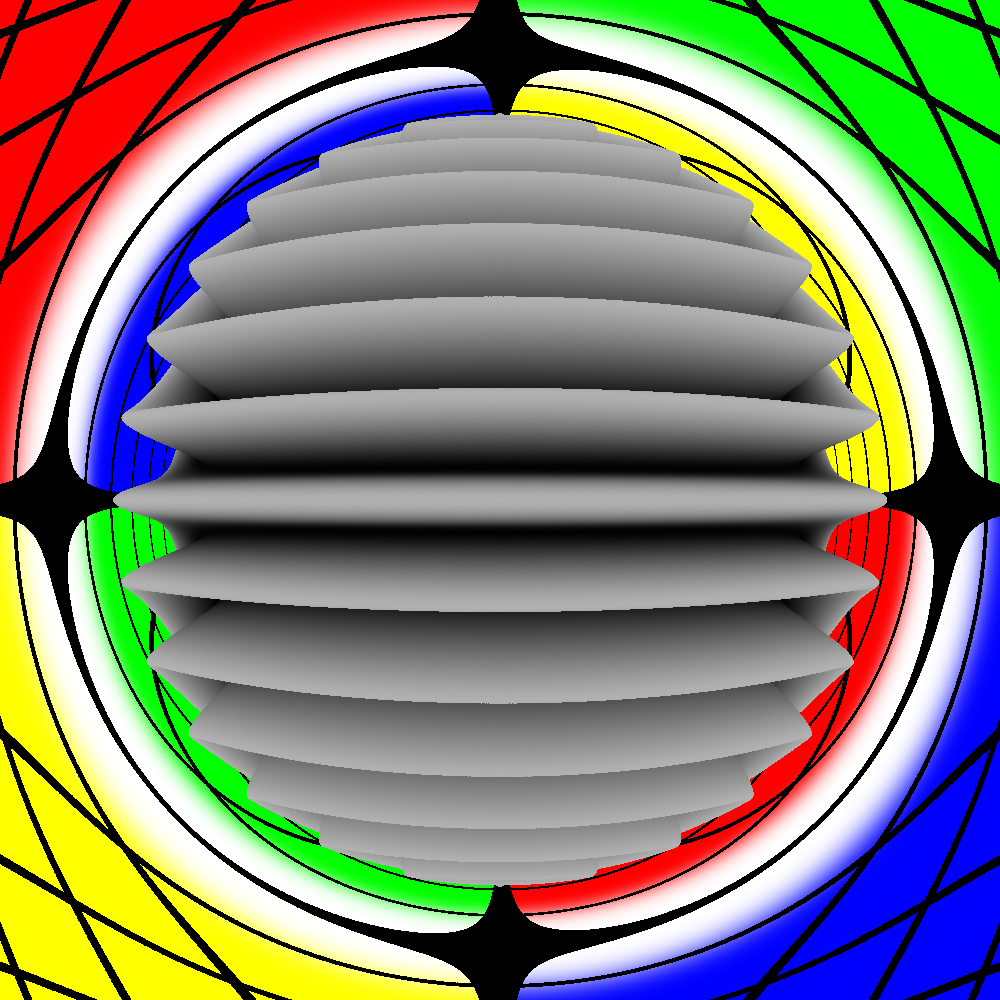}\includegraphics[width=0.4\textwidth]{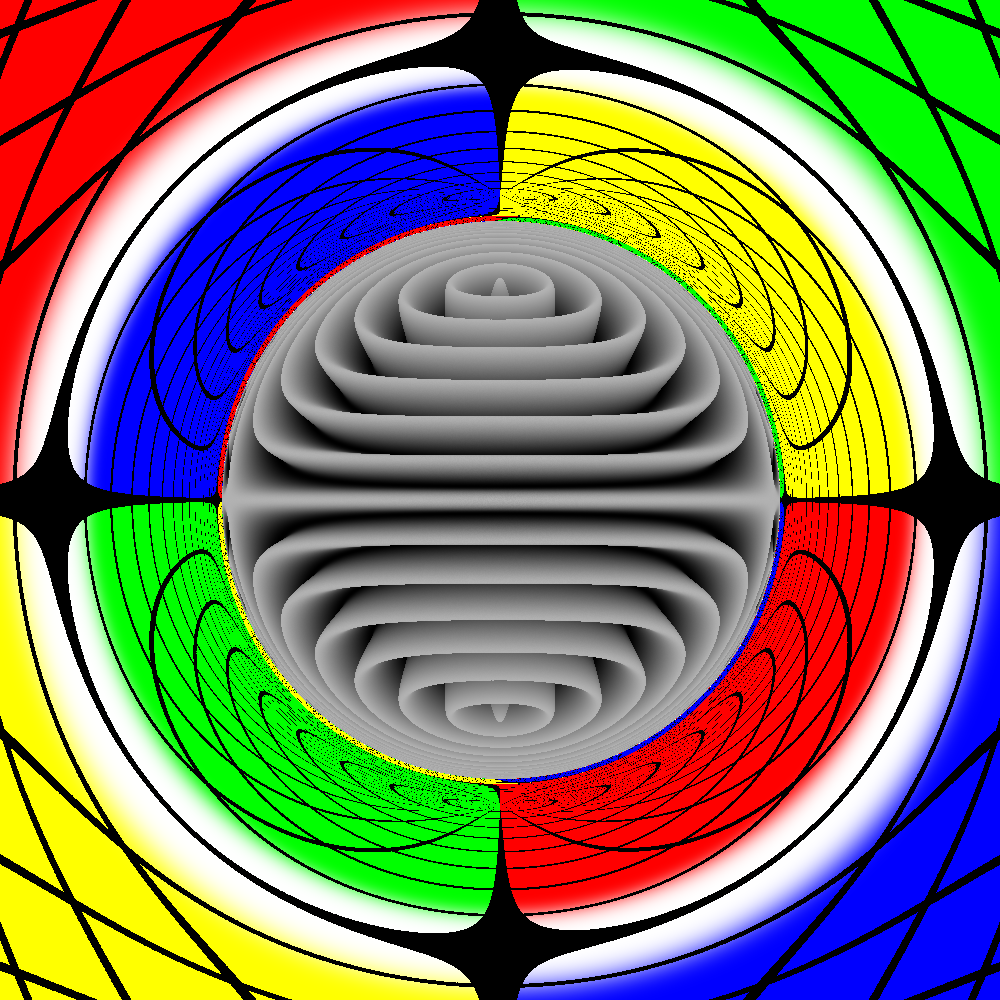}\\
 \includegraphics[width=0.4\textwidth]{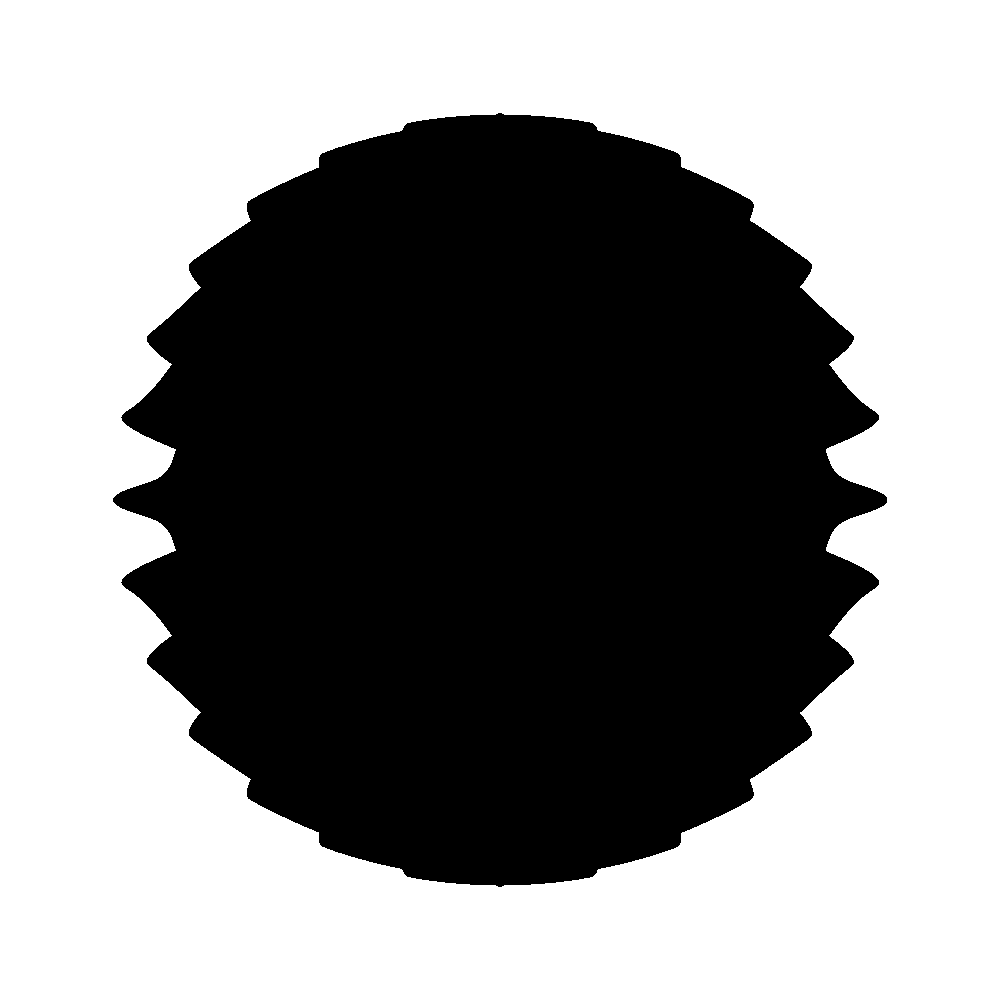}\includegraphics[width=0.4\textwidth]{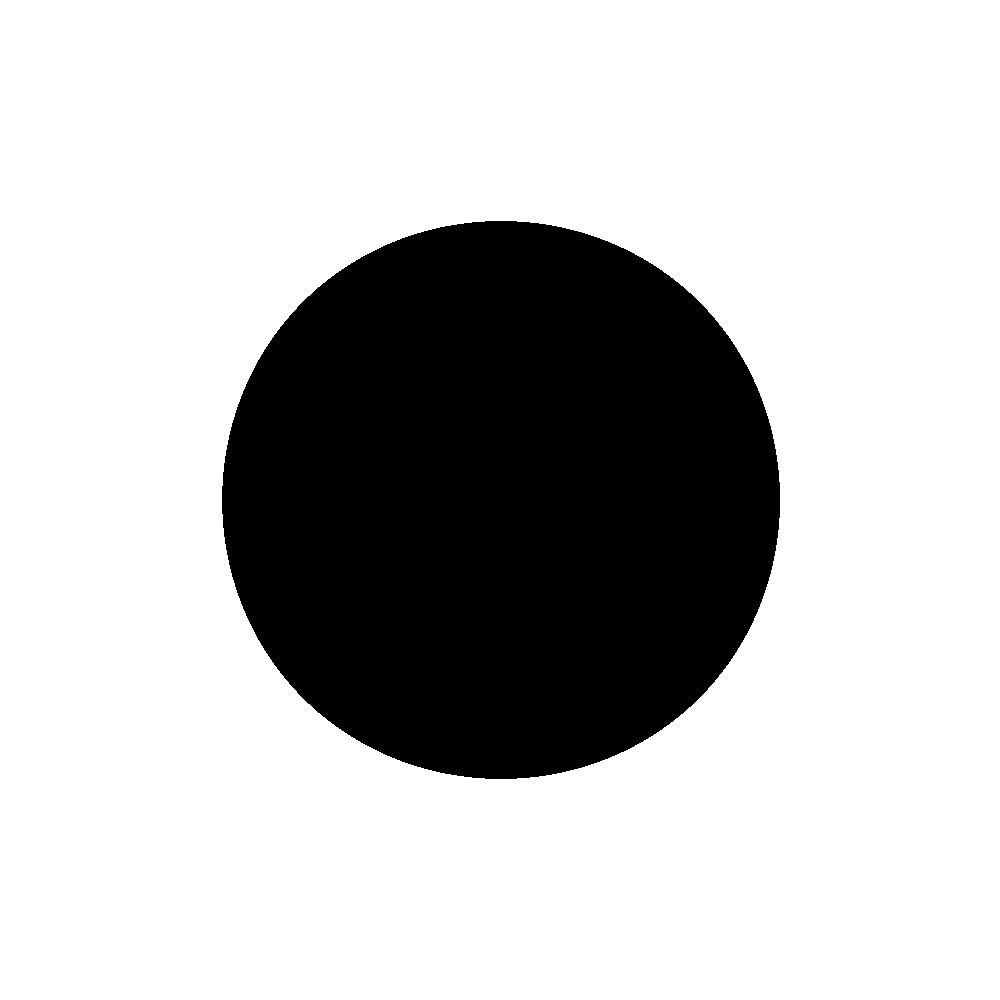}
    \caption{Emitting star in Schwarzschild}
    \end{subfigure}%
    \begin{subfigure}{0.5\textwidth}
      \centering
\includegraphics[width=0.4\textwidth]{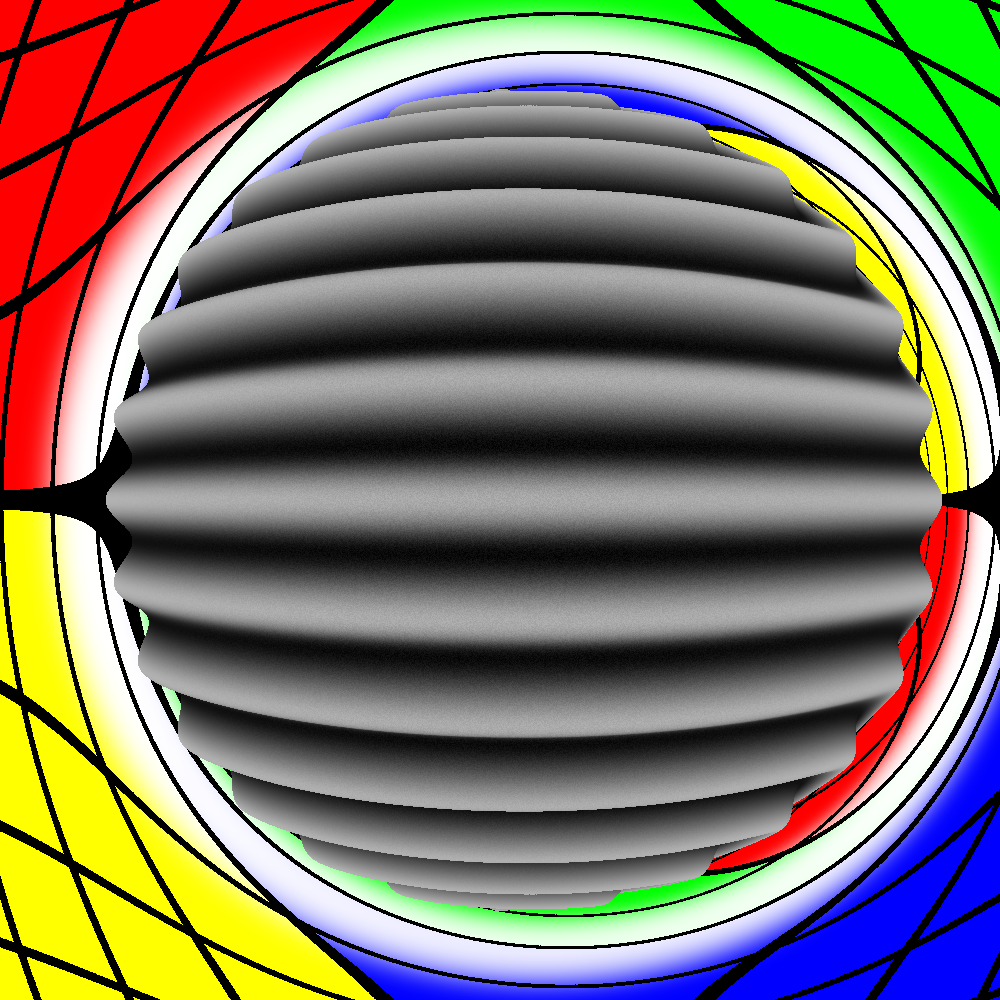}\includegraphics[width=0.4\textwidth]{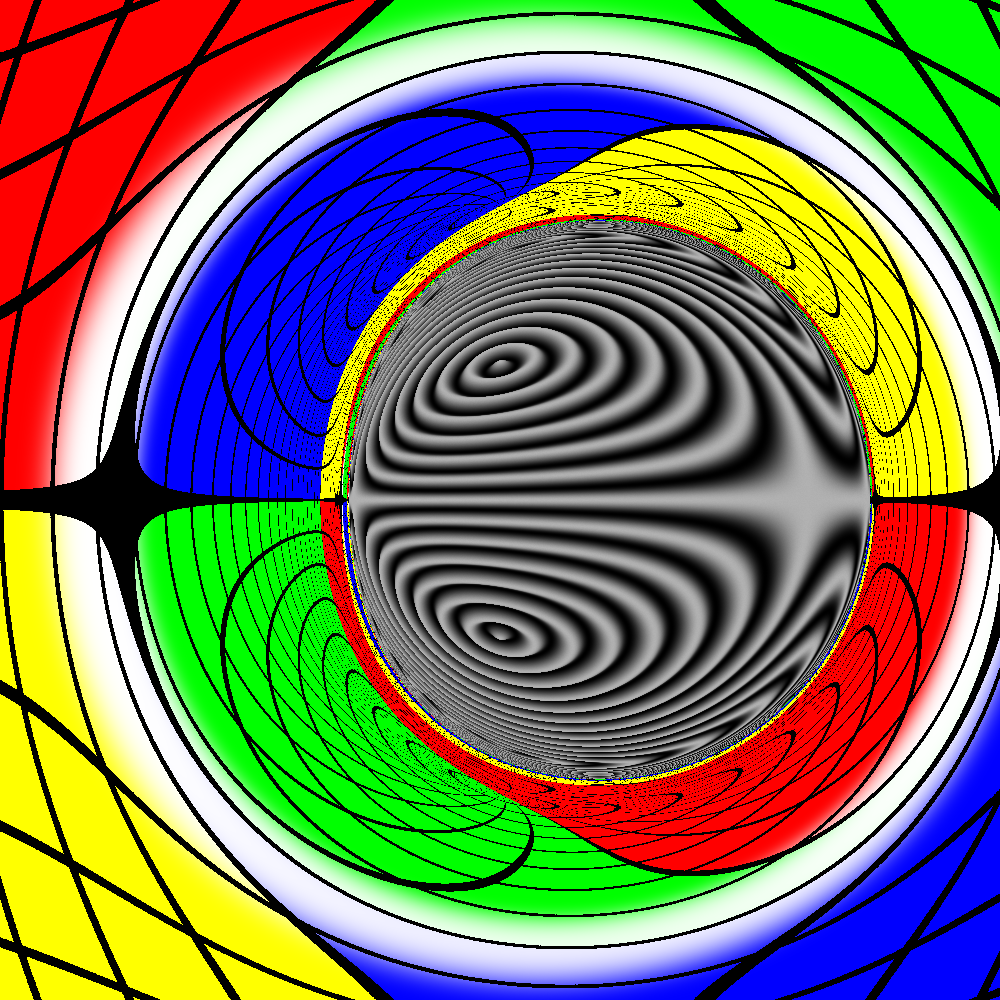}  \\
\includegraphics[width=0.4\textwidth]{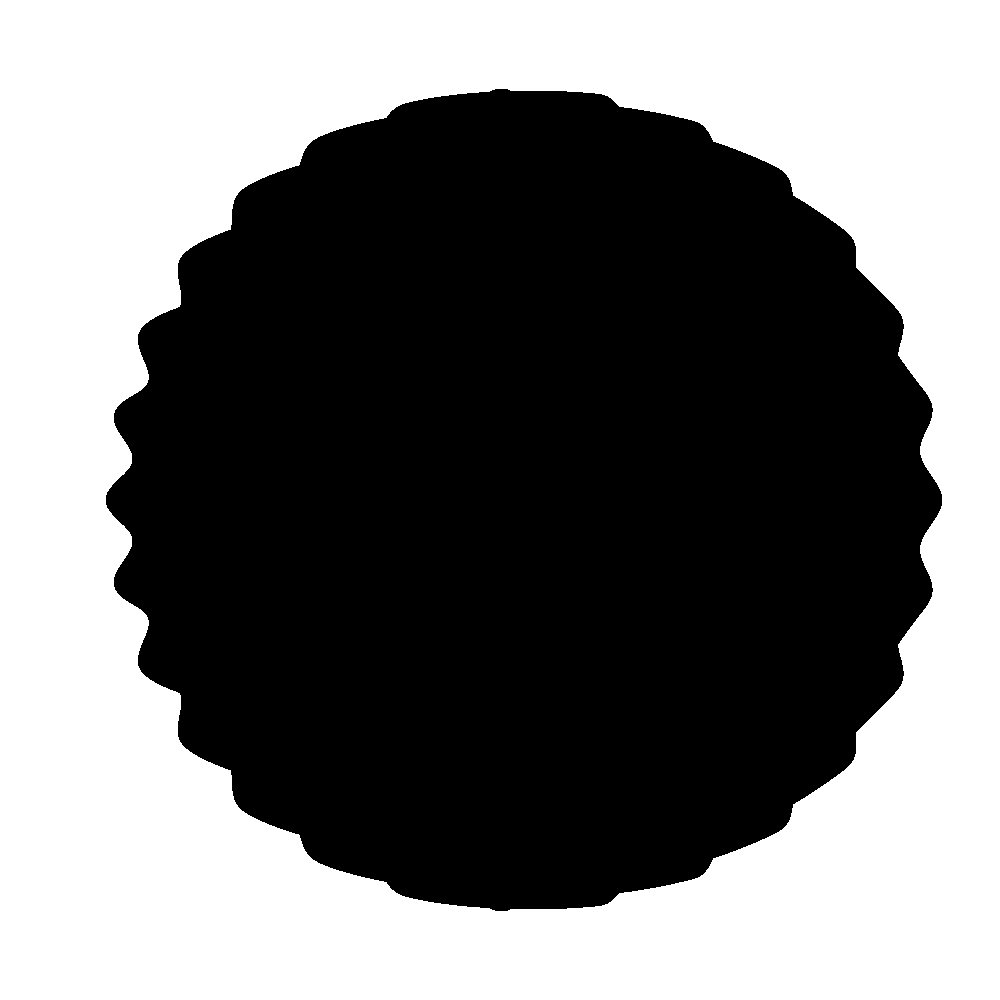}\includegraphics[width=0.4\textwidth]{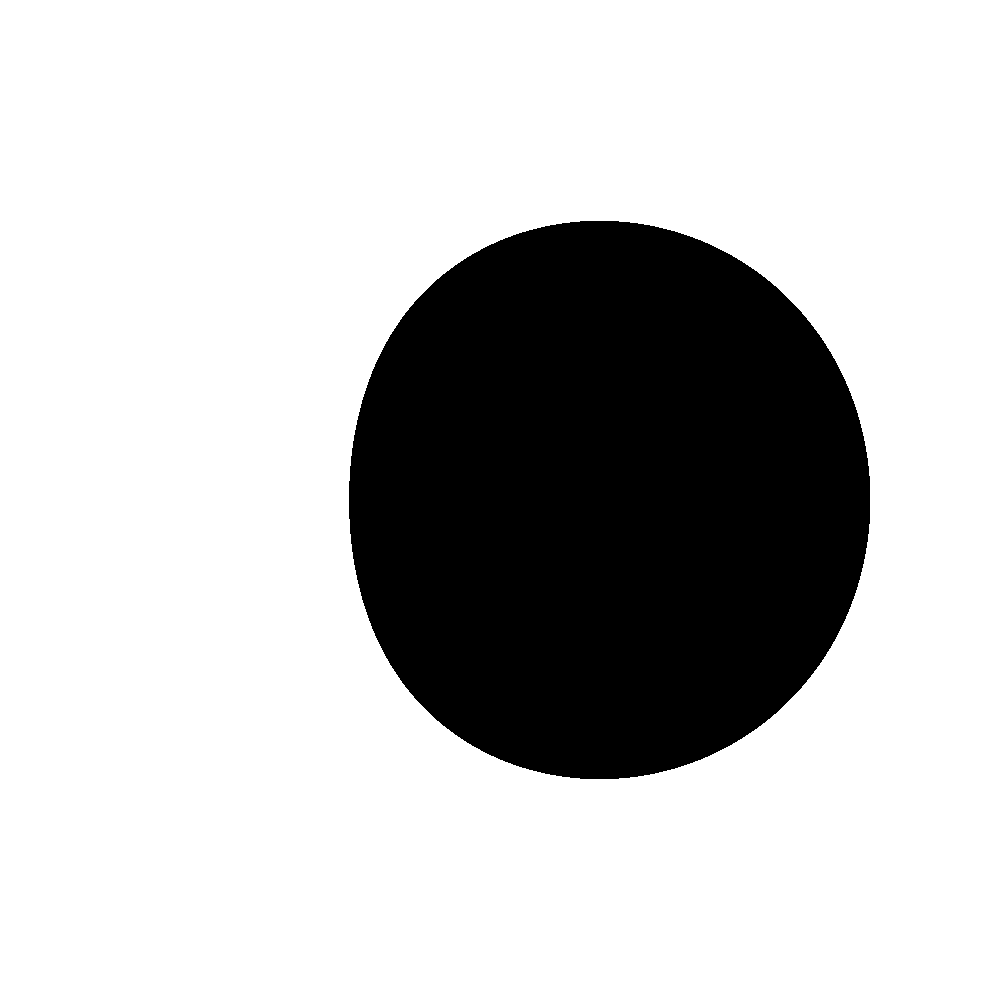}
 \caption{Emitting star in Kerr}
    \end{subfigure}
  \caption {\small {\it Top row $(a)$:} observation images of an emitting star surface in Schwarzschild with $\{r_1,r_2\}\simeq\{4M,\,5.72M\}$ and $\{r_1,r_2\}\simeq\{2.02M,\,2.89M\}$ (from left to right). A darker (brighter) grey color corresponds to a valley (peak) of the star's surface. {\it Bottom row $(a)$:} Silhouette of the previous stars (now totally opaque), with $\mathcal{N}$ painted white. The right image is identical to the Schwarzschild shadow. {\it Top row $(b)$:} observation images of an emitting star surface in Kerr for a rotation parameter $a=0.9M$, with $\{r_1,r_2\}\simeq\{5.74M,\,6.17M\}$ and $\{r_1,r_2\}\simeq\{1.45M,\,1.557M\}$ (from left to right). {\it Bottom row $(b)$:} Similar to $(a)$; the right image is identical to the corresponding Kerr shadow.}
       \label{Reflection-Schw}
\end{figure}

To answer this question, we seek more detail about a bumpy star surface, comprised between two radii $r_1$ and $r_2$:
\[R(\theta)=\left(\frac{r_2-r_1}{2}\right)\cos(28\,\theta) + \left(\frac{r_2+r_1}{2}\right), \qquad R\in[r_1,r_2].\]
placed in a Schwarzschild or Kerr spacetime. Starting in Fig.~\ref{Reflection-Schw} with the Schwarzschild case, the star's wavy structure is clearly visible if $3M<r_1<r_2$ (left column of $(a)$ in Fig.~\ref{Reflection-Schw}). However, when $r_1<r_2\leqslant3M$ (right column of $(a)$ in Fig.~\ref{Reflection-Schw}) the star's outline becomes perfectly circular, and the information from the bumpy surface is lost; had the star been completely opaque and the star's image could not be distinguished from the Schwarzschild shadow, as seen by a far away observer (see bottom row of $(a)$ in Fig.~\ref{Reflection-Schw}). This is a consequence of the photon sphere ($i.e.$ the LR) at $r=3M$, which determines the star's profile in the latter case. In some sense, the shadow is not an image of the horizon but rather that of the FPO structure.

A similar argument applies in Fig.~\ref{Reflection-Schw} in $(b)$, for Kerr with rotation parameter $a=0.9M$. However, the FPO structure now exists in an interval $r\in[r^+,r^-]\simeq [1.558M,\,3.9M]$, where $r^+$ ($r^-$) is the radial coordinate of the co(counter) rotating LR~\cite{Teo2003}. When $r^-<r_1<r_2$, the surface structure is still captured by the star's outline, whereas the latter is identical to the Kerr shadow when $r_1<r_2\leqslant r^+$.

%

%

\section{Closing remarks}
\label{section4}

In this paper we have presented a case study showing that a relevant geometrical feature that produces a deformation of the horizon geometry in a static two BHs system -- the existence of a conical singularity -- leaves no noticeable signature in the corresponding BH shadows. In fact, the shadows of the static two BH system appear similar to those in a dynamical binary, wherein we expect no similar deformation of the BHs' intrinsic geometry. The latter observation suggests further considering light lensing in the static two BH system, or its stationary generalisation -- the double Kerr solution -- with appropriate adaptations, as a proxy of the corresponding process in the corresponding (numerically generated) dynamical binaries.

At the source of the insensitivity of the shadows to the deformations induced by the conical singularity, is the fact that BH shadows are only probing the spacetime geometry as far inside as a set of FPOs, which include LRs, that exist at some distance from the horizon. The shadow is essentially blind to the spacetime region interior to these orbits. We therefore can argue that a deflection $\delta\varphi/2$ in the $\varphi$-direction will produce no net effect for geodesics approaching an FPO since the spatial part of FPOs typically exists on a 2-surface with cylindrical topology, invariant under the action of the Killing vector $\partial_\varphi$.

There is some partial parallelism of this result with the observations in~\cite{Cardoso:2016rao} that the initial ringdown signal emitted by a perturbed ultra-compact object is determined by its LRs structure and it is insensitive to the horizon, even to the extent of its very existence. In this case, however, the later part of the ringdown may yield signatures of the spacetime geometry in the neighborhood of the horizon. 

In the case of lensing we face a situation with some similarities. Two different BHs with a similar FPO structure will cast a similar shadow (see~\cite{Cunha:2016wzk} for an example). And, as illustrated in Section~\ref{section3}, so will \textit{opaque} horizonless ultra-compact objects with a similar FPO structure. Transparent horizonless compact objects, such as boson stars composed of a dark scalar field, on the other hand, can in principle be distinguished~\cite{Cunha:2017wao}. Of course, the potential existence of light sources in between the FPOs and the horizon, could also provide a probe of this spacetime region. Typically, however, light sources are in rapid free fall towards the centre in this region, where there are no long lived orbits.  

Finally, our results on shadows and lensing of the double Schwarzschild BH are also related to geodesic integrability of this solution. In BH solutions wherein geodesic motion is integrable, it has been observed that the BH shadow is always $\mathbb{Z}_2$ symmetric with respect to the image's horizontal axis, even when observed outside the equatorial plane. The lack of such symmetry in the shadows of the double Schwarzschild solution, observed away from the symmetry plane for equal mass BHs, is therefore suggestive that geodesic motion on this background is not Liouville integrable. In other words, no non-trivial Killing tensor exists. Indeed this is the case~\cite{Mirshekari:2010jg}, further supporting the unproved relation between integrability and generic $\mathbb{Z}_2$ symmetry of the shadow.

\section*{Acknowledgements}

We would like to thank E. Radu for discussions. P.C. is supported by Grant No. PD/BD/114071/2015 under the FCT-IDPASC Portugal Ph.D. program. 
C.H. acknowledges funding from the FCT-IF programme. This work was partially supported by the H2020-MSCA-RISE-2015 Grant No. StronGrHEP-690904,
the H2020-MSCA-RISE-2017 Grant No. FunFiCO-777740 and by the CIDMA project UID/MAT/04106/2013. The authors would like to acknowledge networking support by the
COST Action CA16104. The work of MJR was supported by the Max Planck Gesellschaft through the Gravitation and Black Hole Theory Independent Research Group and by NSF grant PHY-1707571 at Utah State University. 



\bibliography{Ref}{}  

\begin{thebibliography}{10}

\bibitem{Dyson:1920cwa}
F.~W. Dyson, A.~S. Eddington, and C.~Davidson, ``{A Determination of the
  Deflection of Light by the Sun's Gravitational Field, from Observations Made
  at the Total Eclipse of May 29, 1919},'' {\em Phil. Trans. Roy. Soc. Lond.},
  vol.~A220, pp.~291--333, 1920.

\bibitem{Shipley:2016omi}
J.~Shipley and S.~R. Dolan, ``{Binary black hole shadows, chaotic scattering
  and the Cantor set},'' {\em Class. Quant. Grav.}, vol.~33, no.~17, p.~175001,
  2016.

\bibitem{Cunha:2017eoe}
P.~V.~P. Cunha, C.~A.~R. Herdeiro, and E.~Radu, ``{Fundamental photon orbits:
  black hole shadows and spacetime instabilities},'' {\em Phys. Rev.},
  vol.~D96, no.~2, p.~024039, 2017.

\bibitem{Cunha:2018acu}
P.~V.~P. Cunha and C.~A.~R. Herdeiro, ``{Shadows and strong gravitational
  lensing: a brief review},'' [1801.00860], 2018.

\bibitem{Cardoso:2014sna}
V.~Cardoso, L.~C.~B. Crispino, C.~F.~B. Macedo, H.~Okawa, and P.~Pani, ``{Light
  rings as observational evidence for event horizons: long-lived modes,
  ergoregions and nonlinear instabilities of ultracompact objects},'' {\em
  Phys. Rev.}, vol.~D90, no.~4, p.~044069, 2014.

\bibitem{Cardoso:2016rao}
V.~Cardoso, E.~Franzin, and P.~Pani, ``{Is the gravitational-wave ringdown a
  probe of the event horizon?},'' {\em Phys. Rev. Lett.}, vol.~116, no.~17,
  p.~171101, 2016.
\newblock [Erratum: Phys. Rev. Lett.117,no.8,089902(2016)].

\bibitem{Cardoso:2016oxy}
V.~Cardoso, S.~Hopper, C.~F.~B. Macedo, C.~Palenzuela, and P.~Pani,
  ``{Gravitational-wave signatures of exotic compact objects and of quantum
  corrections at the horizon scale},'' {\em Phys. Rev.}, vol.~D94, no.~8,
  p.~084031, 2016.

\bibitem{Cardoso:2017cqb}
V.~Cardoso and P.~Pani, ``{Tests for the existence of horizons through
  gravitational wave echoes},'' {\em Nat. Astron.}, vol.~1, pp.~586--591, 2017.

\bibitem{Teo2003}
E.~Teo, ``Spherical photon orbits around a kerr black hole,'' {\em General
  Relativity and Gravitation}, vol.~35, no.~11, pp.~1909--1926, 2003.

\bibitem{Cunha:2015yba}
P.~V.~P. Cunha, C.~A.~R. Herdeiro, E.~Radu, and H.~F. Runarsson, ``{Shadows of
  Kerr black holes with scalar hair},'' {\em Phys. Rev. Lett.}, vol.~115,
  no.~21, p.~211102, 2015.

\bibitem{Dolan:2016bxj}
S.~R. Dolan and J.~O. Shipley, ``{Stable photon orbits in stationary
  axisymmetric electrovacuum spacetimes},'' {\em Phys. Rev.}, vol.~D94, no.~4,
  p.~044038, 2016.

\bibitem{Cunha:2016bjh}
P.~V.~P. Cunha, J.~Grover, C.~Herdeiro, E.~Radu, H.~Runarsson, and A.~Wittig,
  ``{Chaotic lensing around boson stars and Kerr black holes with scalar
  hair},'' {\em Phys. Rev.}, vol.~D94, no.~10, p.~104023, 2016.

\bibitem{Cunha:2017wao}
P.~V.~P. Cunha, J.~A. Font, C.~Herdeiro, E.~Radu, N.~Sanchis-Gual, and
  M.~Zilh{\~a}o, ``{Lensing and dynamics of ultracompact bosonic stars},'' {\em
  Phys. Rev.}, vol.~D96, no.~10, p.~104040, 2017.

\bibitem{Cunha:2017qtt}
P.~V.~P. Cunha, E.~Berti, and C.~A.~R. Herdeiro, ``{Light ring stability in
  ultra-compact objects},'' {\em Phys. Rev. Lett.}, vol.~119, no.~25,
  p.~251102, 2017.

\bibitem{Bardeen1973}
J.~M. Bardeen, {\em Timelike and null geodesies in the Kerr metric}.
\newblock C. Witt and B. Witt, editors, Black Holes, pp.215, 1973.

\bibitem{Falcke:1999pj}
H.~Falcke, F.~Melia, and E.~Agol, ``{Viewing the shadow of the black hole at
  the galactic center},'' 1999.

\bibitem{ZAKHAROV2005479}
A.~Zakharov, A.~Nucita, F.~DePaolis, and G.~Ingrosso, ``Measuring the black
  hole parameters in the galactic center with radioastron,'' {\em New
  Astronomy}, vol.~10, no.~6, pp.~479 -- 489, 2005.

\bibitem{Psaltis:2014mca}
D.~Psaltis, F.~Ozel, C.-K. Chan, and D.~P. Marrone, ``{A General Relativistic
  Null Hypothesis Test with Event Horizon Telescope Observations of the
  black-hole shadow in Sgr A*},'' {\em Astrophys. J.}, vol.~814, no.~2, p.~115,
  2015.

\bibitem{Bambi:2015rda}
C.~Bambi, ``{Testing the Kerr Paradigm with the Black Hole Shadow},'' in {\em
  {Proceedings, 14th Marcel Grossmann Meeting on Recent Developments in
  Theoretical and Experimental General Relativity, Astrophysics, and
  Relativistic Field Theories (MG14) (In 4 Volumes): Rome, Italy, July 12-18,
  2015}}, vol.~4, pp.~3494--3499, 2017.

\bibitem{Johannsen:2015hib}
T.~Johannsen, A.~E. Broderick, P.~M. Plewa, S.~Chatzopoulos, S.~S. Doeleman,
  F.~Eisenhauer, V.~L. Fish, R.~Genzel, O.~Gerhard, and M.~D. Johnson,
  ``{Testing General Relativity with the Shadow Size of Sgr A*},'' {\em Phys.
  Rev. Lett.}, vol.~116, no.~3, p.~031101, 2016.

\bibitem{Vincent:2015xta}
F.~H. Vincent, Z.~Meliani, P.~Grandclement, E.~Gourgoulhon, and O.~Straub,
  ``{Imaging a boson star at the Galactic center},'' {\em Class. Quant. Grav.},
  vol.~33, no.~10, p.~105015, 2016.

\bibitem{Mars:2017jkk}
M.~Mars, C.~F. Paganini, and M.~A. Oancea, ``{The fingerprints of black
  holes-shadows and their degeneracies},'' {\em Class. Quant. Grav.}, vol.~35,
  no.~2, p.~025005, 2018.

\bibitem{Bach:1922}
R.~Bach and H.~Weyl, ``{Neue l\"osungen der Einsteinschen
  gravitationsgleichungen},'' {\em Math. Zeits.}, vol.~13, pp.~134--145, 1922.

\bibitem{Weyl:1917gp}
H.~Weyl, ``{The theory of gravitation},'' {\em Annalen Phys.}, vol.~54,
  pp.~117--145, 1917.

\bibitem{Einstein:1936fp}
A.~Einstein and N.~Rosen, ``{Two-Body Problem in General Relativity Theory},''
  {\em Phys. Rev.}, vol.~49, pp.~404--405, 1936.

\bibitem{Israel:1964}
W.~Israel and K.~A. Khan, ``{Colinear particles and Bondi dipoles in General
  Relativity},'' {\em Nouvo Cimento}, vol.~33, p.~331, 1964.

\bibitem{Costa:2000kf}
M.~S. Costa and M.~J. Perry, ``{Interacting black holes},'' {\em Nucl. Phys.},
  vol.~B591, pp.~469--487, 2000.

\bibitem{Bohn:2014xxa}
A.~Bohn, W.~Throwe, F.~H{\'e}bert, K.~Henriksson, D.~Bunandar, M.~A. Scheel,
  and N.~W. Taylor, ``{What does a binary black hole merger look like?},'' {\em
  Class. Quant. Grav.}, vol.~32, no.~6, p.~065002, 2015.

\bibitem{Grenzebach:2015oea}
A.~Grenzebach, V.~Perlick, and C.~Lämmerzahl, ``{Photon Regions and Shadows of
  Accelerated Black Holes},'' {\em Int. J. Mod. Phys.}, vol.~D24, no.~09,
  p.~1542024, 2015.

\bibitem{PhysRevD.80.104036}
F.~S. Coelho and C.~A.~R. Herdeiro, ``Relativistic euler's three-body problem,
  optical geometry, and the golden ratio,'' {\em Phys. Rev. D}, vol.~80,
  p.~104036, Nov 2009.

\bibitem{Yumoto:2012kz}
A.~Yumoto, D.~Nitta, T.~Chiba, and N.~Sugiyama, ``{Shadows of Multi-Black
  Holes: Analytic Exploration},'' {\em Phys. Rev.}, vol.~D86, p.~103001, 2012.

\bibitem{Abdolrahimi:2015rua}
S.~Abdolrahimi, R.~B. Mann, and C.~Tzounis, ``{Distorted Local Shadows},'' {\em
  Phys. Rev.}, vol.~D91, no.~8, p.~084052, 2015.

\bibitem{Costa:2009wj}
M.~S. Costa, C.~A.~R. Herdeiro, and C.~Rebelo, ``{Dynamical and Thermodynamical
  Aspects of Interacting Kerr Black Holes},'' {\em Phys. Rev.}, vol.~D79,
  p.~123508, 2009.

\bibitem{Gibbons:2009qe}
G.~W. Gibbons, C.~A.~R. Herdeiro, and C.~Rebelo, ``{Global embedding of the
  Kerr black hole event horizon into hyperbolic 3-space},'' {\em Phys. Rev.},
  vol.~D80, p.~044014, 2009.

\bibitem{Cunha:2016wzk}
P.~V.~P. Cunha, C.~A.~R. Herdeiro, B.~Kleihaus, J.~Kunz, and E.~Radu,
  ``{Shadows of Einstein-dilaton-Gauss-Bonnet black holes},'' {\em Phys.
  Lett.}, vol.~B768, pp.~373--379, 2017.

\bibitem{Mirshekari:2010jg}
S.~Mirshekari and C.~M. Will, ``{Carter-like constants of motion in the
  Newtonian and relativistic two-center problems},'' {\em Class. Quant. Grav.},
  vol.~27, p.~235021, 2010.

\end{thebibliography}
\bibliographystyle{ieeetr}

\end{document}